\documentclass[numberedappendix,onecolumn]{emulateapj} 
\usepackage{rotating} 
\usepackage{amsmath,amssymb,bbold,mathrsfs}
\usepackage{subfigure}
\usepackage{bm}

\begin{document}

\shorttitle{Forward-scattering polarization in \ion{Sr}{1} 4607~\AA\ with limited time resolution}
\title{The impact of limited time resolution on the \\forward-scattering polarization 
in the solar \ion{Sr}{1} 4607~\AA\ line}

\author{T.\ del Pino Alem\'an$^{1,2}$, J.\ Trujillo Bueno$^{1,2,3}$\footnote{Affiliate scientist 
of the National Center for Atmospheric Research, Boulder, U.S.A.}}

\affil{$^1$Instituto de Astrof\'{\i}sica de Canarias, 38205, La Laguna, Tenerife, Spain}
\affil{$^2$Departamento de Astrof\'\i sica, Facultad de F\'\i sica, Universidad de La Laguna,
Tenerife, Spain}
\affil{$^3$Consejo Superior de Investigaciones Cient\'{\i}ficas, Spain}

\begin{abstract}

Theoretical investigations predicted that high spatio-temporal resolution observations in
the \ion{Sr}{1} 4607~\AA\ line must show a conspicuous scattering polarization pattern at
the solar disk-center, which encodes information on the unresolved magnetism of the
inter-granular photospheric plasma. Here we present a study of the impact of limited time
resolution on the observability of such forward scattering (disk-center) polarization signals.
Our investigation is based on three-dimensional radiative transfer calculations in a
time-dependent magneto-convection model of the quiet solar photosphere, taking into account
anisotropic radiation pumping and the Hanle effect. This type of radiative transfer simulation
is computationally costly, reason why the time variation had not been investigated before
for this spectral line. We compare our theoretical results with recent disk-center filter
polarimetric observations in the \ion{Sr}{1} 4607~\AA\ line, showing that there is good
agreement in the polarization patterns. We also show what we can expect to observe
with the Visible Spectro-Polarimeter at the upcoming Daniel K. Inouye Solar Telescope.

\end{abstract}

\keywords{Polarization - scattering - radiative transfer - Sun: photosphere -
          Sun: magnetism}


\section{Introduction}\label{S-intro}

Even the quietest regions of the solar atmosphere are permeated by magnetic
activity, happening at scales smaller than what we have been able to
historically resolve. Recent investigations indicate that the magnetic energy
density stored at these scales is so significant that this small-scale
magnetism could potentially be the main driver for heating the outer layers
of the solar atmosphere above the quiet regions of the solar disk
(\citealt{Trujilloetal2004,Amarietal2015,Rempel2017}).

The radiation field within the solar atmosphere is anisotropic because 
there is an unbalance between the radiation traveling upwards and sideways.
As a result, the radiatively-induced transitions produce atomic level
polarization, which in turns produces the so-called scattering line
polarization. This occurs even in the absence of magnetic fields. 
If the intensity of the pumping radiation had axial symmetry around the
local vertical, the only source of scattering polarization at the solar 
disk-center would be the presence of a non-vertical magnetic field
\citep{Trujillo2001,BLandiLandolfi2004}. However,
the situation in the Sun is more interesting, as both the horizontal
inhomogeneities and velocity gradients break the axial symmetry 
of the incident radiation field, producing scattering polarization at
disk-center even in the absence of magnetic fields
(e.g., \citealt{MansoTrujillo2011,StepanTrujillo2016}). Therefore, with
some exceptions (e.g., \citealt{Trujilloetal2002}), the observation of
forward scattering polarization is not indicative of magnetic activity by
itself. However, the degree and angle of this polarization greatly depends
on the magnetic field vector in the region of formation of the spectral
line under consideration.

One spectral line of especial interest to study the small-scale magnetic
activity in the quiet regions of the solar photosphere is the \ion{Sr}{1}
transition at 4607~\AA\ , because it shows polarization amplitudes among
the largest in the second solar spectrum (\citealt{Stenfloetal1997,BGandorfer2002}),
its scattering polarization is sensitive to magnetic field strengths as
large as 300~G, and because its polarization can be reliably calculated
using a two-level atomic model. In
\citeyear{Trujilloetal2004}, advanced theoretical modeling of
spectropolarimetric data for this spectral line led to the discovery of a
vast amount of hidden magnetic energy in the quiet Sun, due to the
presence of a small-scale magnetic field with a mean field strength of the
order of $\sim100$~G (\citealt{Trujilloetal2004}), concluding that the
statistical properties of the photospheric magnetic field in the
quiet-Sun inter-network regions vary at the spatial scales of the solar
granulation pattern.

Recently, \cite{delPinoetal2018} solved the radiation transfer problem of
scattering polarization in the \ion{Sr}{1} 4607~\AA\ line using a high-resolution
three-dimensional (3D) model resulting from advanced magneto-convection
simulations \citep{Rempel2014}. The selected snapshot model has a
mean field strength of $170$~G at the model's visible surface and a convection
zone magnetized close to the equipartition. They demonstrated that the magnetic
field of the model's photosphere produces a Hanle depolarization compatible with
scattering polarization observations of the \ion{Sr}{1} 4607~\AA\ line
without spatio-temporal resolution. The
authors continued and expanded the investigation of \cite{TrujilloShchukina2007}
providing useful information to facilitate high spatio-temporal resolution
observational advances, in particular
the detection of the theoretically-predicted disk-center polarization patterns.

Motivated by the theoretical investigation of \cite{TrujilloShchukina2007},  
observations with the Z\"urich Imaging Polarimeter (ZIMPOL) have been  
carried out at the GREGOR telescope of the Observatorio del Teide 
(Tenerife; Spain) to study the spatial variability of the polarization signals
at different distances from the solar limb (\citealt{Biandaetal2018,Dharaetal2019}),
finding variations at granular scales for all the observed heliocentric angles. 
More recently, \cite{Zeuneretal2020} used the Fast Solar Polarimeter
(FSP 2) attached to the Dunn Solar Telescope of the Sacramento Peak
Observatory (USA) and applied a novel strategy to increase the signal to noise
ratio of their disk-center filter polarimetric observations in the \ion{Sr}{1}
4607~\AA\ line. Although their successful detection of forward scattering polarization 
confirms the theoretical predictions, a satisfactory confrontation with the recent  
theoretical results of \cite{delPinoetal2018} requires taking into account 
the instrumental degradation produced by their specific observational setup.

The solution of the radiative transfer problem of scattering polarization in   
3D models of the solar atmosphere is very computationally intensive.
For this reason, the previous investigations used a single snapshot
from a hydrodynamical or magneto-hydrodynamical simulation, neglecting the
impact on the polarization signals due to the limited temporal resolution.
In this work we extend our previous study (\citealt{delPinoetal2018}) by
including the effect of the temporal evolution on the observed scattering
polarization in a 3D time-dependent model of the solar photosphere.
To this end, we have solved the aforementioned radiative transfer
problem in 151 3D snapshots of a magneto-convection simulation
covering five minutes of solar time. In section \S\ref{Sproblem} we summarize
the properties of the time series we use and the synthesis method. In section
\S\ref{Stime} we study the effect of the limited time resolution on the emergent
Stokes profiles. Finally, in section \S\ref{Sobs} we study the joint effect of
the limited time resolution and the instrumental effects for two observational
setups of interest: the filter polarimeter used by \cite{Zeuneretal2020}
and the (slit-based) Visible Spectro-Polarimeter (ViSP) attached to the Daniel
K. Inouye Solar Telescope (DKIST).

\section{The Physical Problem}\label{Sproblem}

The 3D model of the quiet solar photosphere used in this investigation is a time
series of a magneto-convection simulation by \cite{Rempel2014}.
The original grid of the 3D snapshots has
$384\times 384\times 256$ points, with a regular spacing of $16$~km in the three
dimensions. For our calculations, we have cut the magneto-hydrodynamical (MHD)
model in the vertical direction in order to include only the region of the
atmosphere that is relevant for the formation of the \ion{Sr}{1} 4607~\AA\ line.
The atmospheric model we use in our calculations thus has
$384\times 384\times 108$ grid points, with the vertical axis going from
$\sim850$~km below to $\sim850$~km above the average height where the optical
depth of the continuum at 4607~\AA\ is unity, and the same $16$~km grid
resolution. The simulation had a fixed time step of $0.20625$~s, storing one
every ten time steps; therefore, the snapshots have a cadence of $2.0625$~s.

The calculations of the emergent Stokes profiles have been carried out as in
\cite{delPinoetal2018}, with the radiative transfer code PORTA
\citep[see][]{StepanTrujillo2013}\footnote{PORTA is publicly available at
{\url https://gitlab.com/polmag/PORTA}}, which solves the non-LTE multilevel
problem of the generation and transfer of polarized radiation in 3D Cartesian
models of stellar atmospheres taking fully into account the breaking of the
axial symmetry of the incident radiation field at each point within the medium.

We first solved the problem of the strontium ionization balance in order to
obtain the number density of strontium atoms in the lower and upper levels of
the \ion{Sr}{1} line at 4607~\AA\ at each spatial point of the 3D model. To this
end, we used the same 15 levels atomic model and abundance than in
\cite{delPinoetal2018} and the same numerical code than in
\cite{delPinoetal2020}. We then used a two-level atomic model to compute
with PORTA the Stokes profiles of the \ion{Sr}{1} 4607~\AA\ line.
For the rates of depolarizing collisions with neutral hydrogen atoms
we have taken the expression given by \cite{Faurobertetal1995}. 

In our 3D radiative transfer calculations of the scattering polarization
in the \ion{Sr}{1} 4607~\AA\ line we do not include the polarization of the
continuum radiation caused by Rayleigh and Thomson scattering
\citep[see][]{TrujilloShchukina2009}. This is a suitable approximation because
at 4607~\AA\ the continuum polarization amplitude is, in general, much smaller
than that of the line itself. 

\begin{figure}[htp]
\centering 
\includegraphics[width=.45\textwidth]{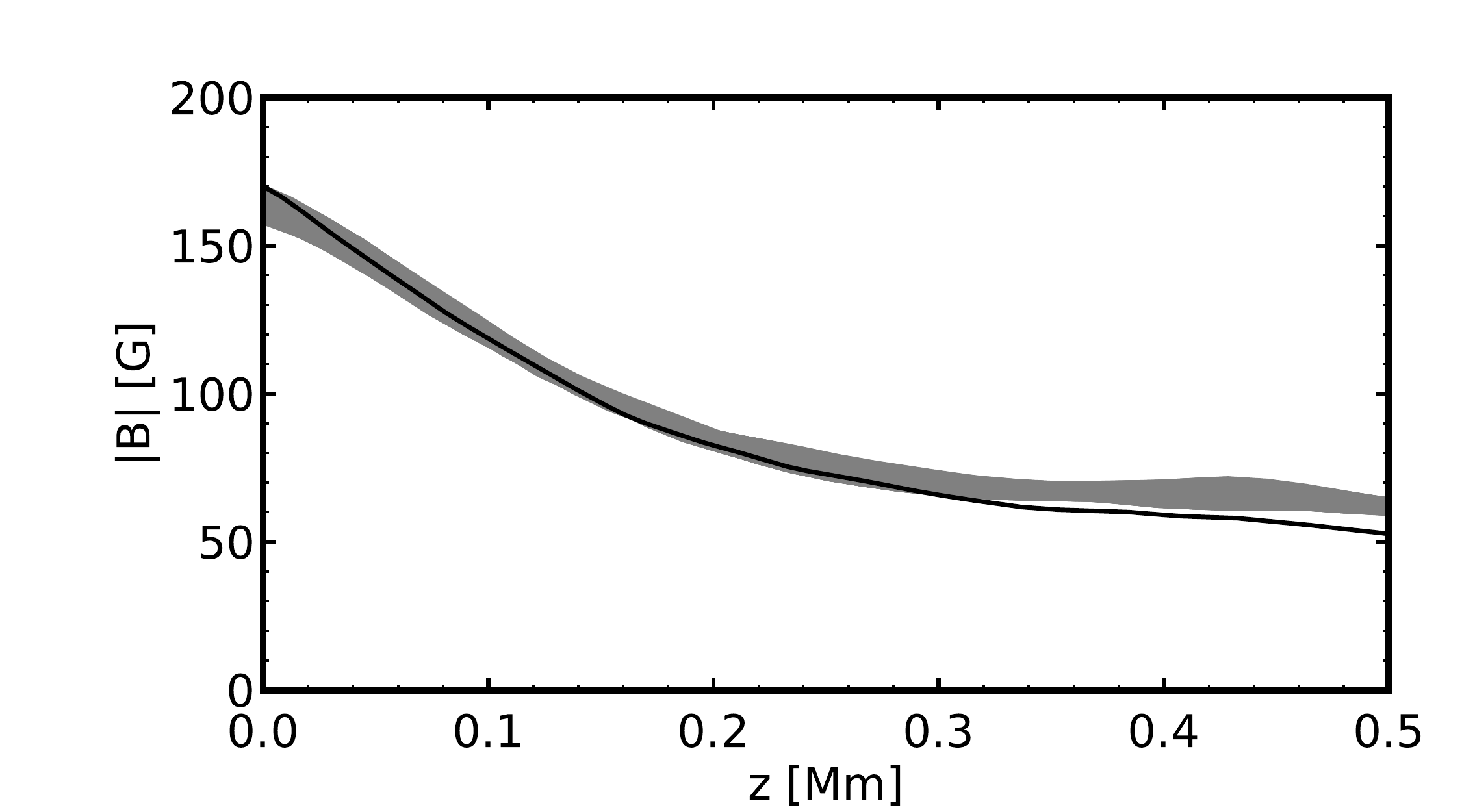}
\includegraphics[width=.45\textwidth]{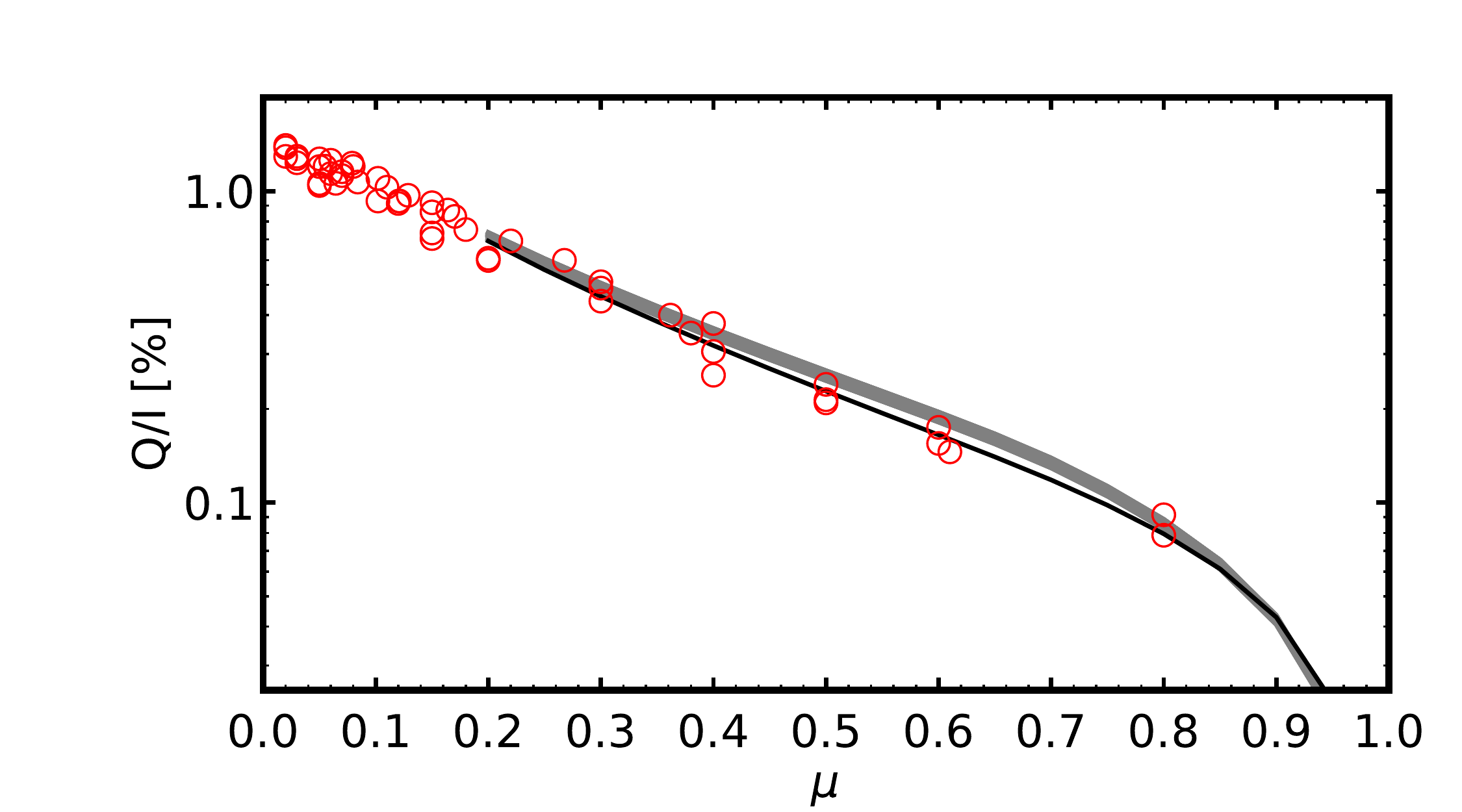}
\caption{{\bf Left panel}: Variation with height of the horizontal average of
the magnetic field strength for all the snapshots in the time series (gray
curves) and the 3D snapshot model used in \cite{delPinoetal2018} (black curve).
{\bf Right panel}: Center-to-limb variation of the spatially and azimuthally
averaged fractional linear polarization for all the snapshots in the time
series (gray curves) and the 3D snapshot model used in \cite{delPinoetal2018}
(black curve). The symbols correspond to the observations considered by
\cite{Trujilloetal2004}, taken during a minimum and maximum of the solar
activity cycle.}
\label{F-2018compare} 
\end{figure}

The time series studied in this paper ($16$~km spatial resolution) is not
the same one from where the snapshot used by
\cite{delPinoetal2018} was extracted (8~km spatial resolution).
The left panel in Fig. \ref{F-2018compare} shows the variation with height of
the average magnetic field strength in all the $16$~km snapshots (gray curves)
and in the $8$~km model used in \cite{delPinoetal2018} (black curve); they are very
similar in the region of formation of the \ion{Sr}{1} line at 4607~\AA\
(approximately between 170 and 350~km above the visible continuum surface)
and the 8~km one shows an average magnetic field strength within the range of
variability of the time series.
The right panel in Fig. \ref{F-2018compare} shows the CLV of the average
fractional linear polarization in all the snapshots (gray curves) and in the single
snapshot model used by \cite{delPinoetal2018} (black curve), while the red circles
show various observations taken during a minimum and maximum of the solar activity
cycle (see \citealt{Trujilloetal2004}). The CLV obtained from the high-resolution
(8~km) single snapshot model (black curve) produces an excellent fit to the
observations. The fit provided by the CLV obtained from the lower-resolution (16~km)
time series models is also satisfactory, although it is not as extraordinary as that
provided by the 8~km magneto-convection model. Note that, regardless of the small
differences seen in Fig. \ref{F-2018compare}, both CLV tend to zero at disk-center, 
as the plotted $Q/I$ quantity lacks any spatial resolution. In summary, the time
series snapshot models resulting from Rempel's (2014) 16~km resolution
magneto-convection simulations are fully appropriate for this investigation and,
in particular, for investigating the relative impact of the time resolution on the
scattering polarization.

\section{The Effect of Time Resolution}\label{Stime}

In this section we study the effect that a limited time resolution has on the
spatial pattern and amplitudes of the linear polarization signals for
observations at the disk-center. The solar atmosphere is continuously evolving,
so it varies during the exposure time of our observations. Because
the polarization is a signed quantity, it is significantly more affected by this
evolution than the intensity. We are interested, in particular, on how the
exposure time affects the mean amplitude of the line center polarization of our
synthetic profiles, as well as the effect on the fine spatial structure and
fluctuation pattern.

\begin{figure}[htp]
\centering 
\includegraphics[width=.95\textwidth]{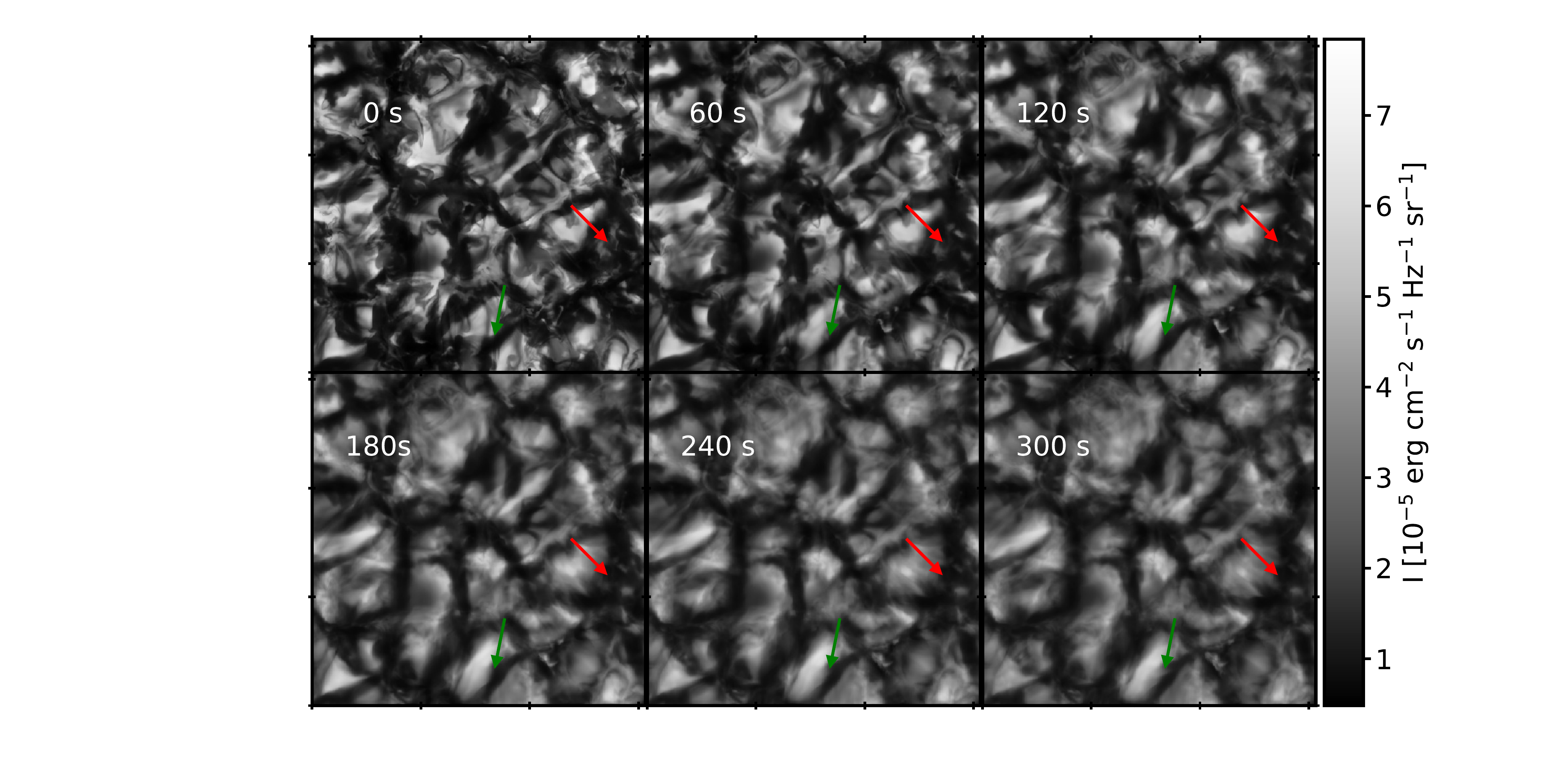} \\\vspace*{-2.85em}
\includegraphics[width=.95\textwidth]{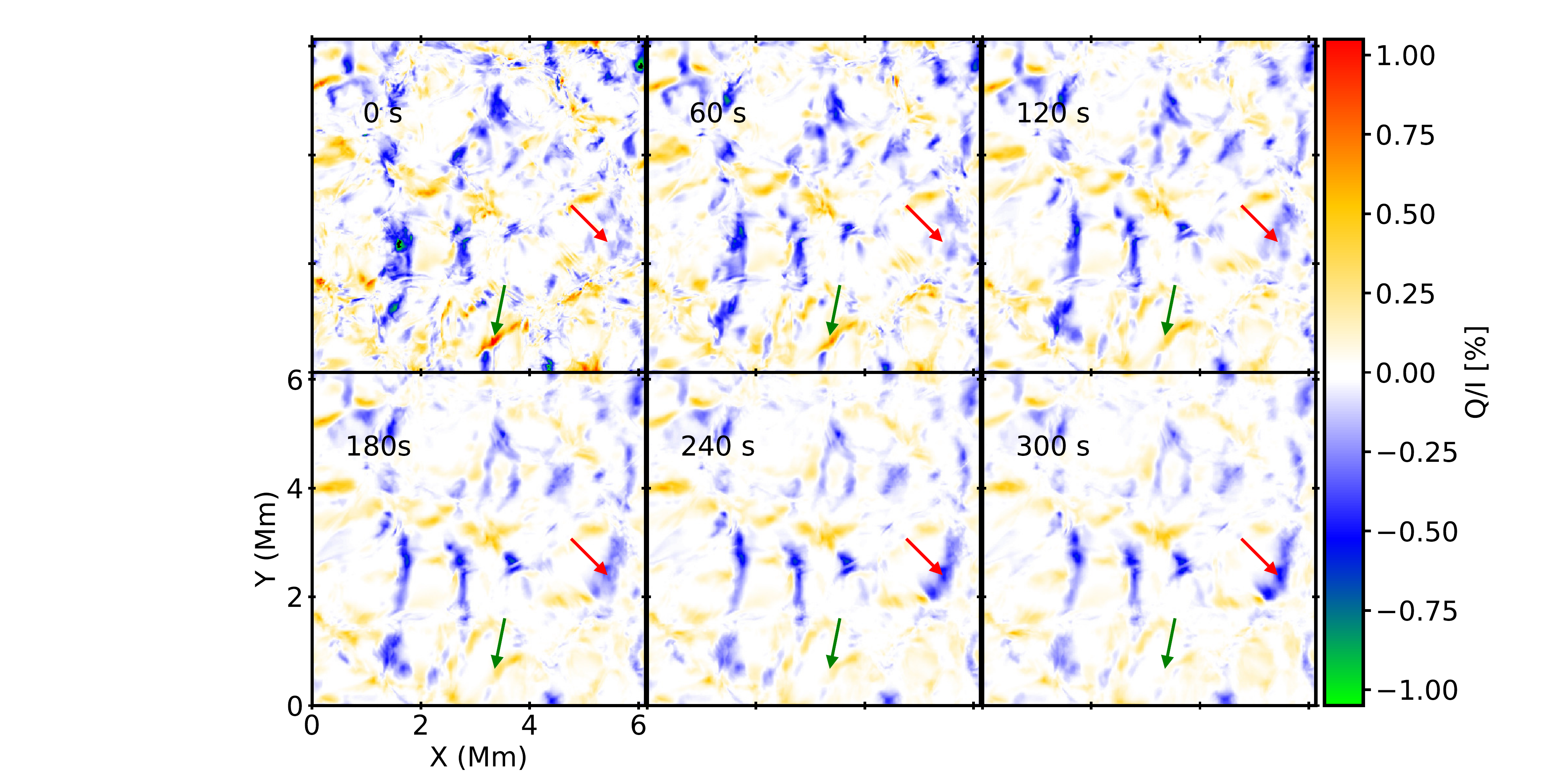}
\caption{Intensity (top panels) and fractional linear polarization $Q/I$
(bottom panels) at the \ion{Sr}{1} 4607~\AA\ line center, for the disk-center
LOS, for the first snapshot of the time series and for different integration
times, from top to bottom and left to right: $0\,$s, $60\,$s, $120\,$s,
$180\,$s, $240\,$s, and $300\,$s.}
\label{F-timeintegrals} 
\end{figure}

In order to study this effect, we integrate the Stokes parameters for each
wavelength and for each point in the field of view among a number of snapshots
covering different exposure times. Figure \ref{F-timeintegrals} shows the
intensity and the fractional linear polarization $Q/I$ at the line center for
the disk-center line of sight, for the first snapshot of the time series and for
different exposure times; from left to right and top to bottom,
{\it instantaneous}, $60$~s, $120$~s, $180$~s, $240$~s, and $300$~s. As expected,
the impact of the exposure time on the scattering polarization is apparent, at
plot level, for smaller times than for the intensity. The intensity is blurred
and gradually loses contrast, very noticeable especially in the bottom row of
the intensity panels in Fig. \ref{F-timeintegrals} (integration time $>120$~s).
Regarding the fractional linear polarization $Q/I$, the loose of details is
already easily seen for $60$~s, and the decrease of contrast is more apparent
due to the signed nature of the polarization. Notice that, while in most of the
field of view the time integration results in decreasing signals (see green
arrow in Fig. \ref{F-timeintegrals} for an example) there are some regions
where the polarization increases as a result of the time evolution of the
granulation pattern (e.g., red arrow in Fig. \ref{F-timeintegrals}).

\begin{figure}[htp]
\centering 
\includegraphics[width=.45\textwidth]{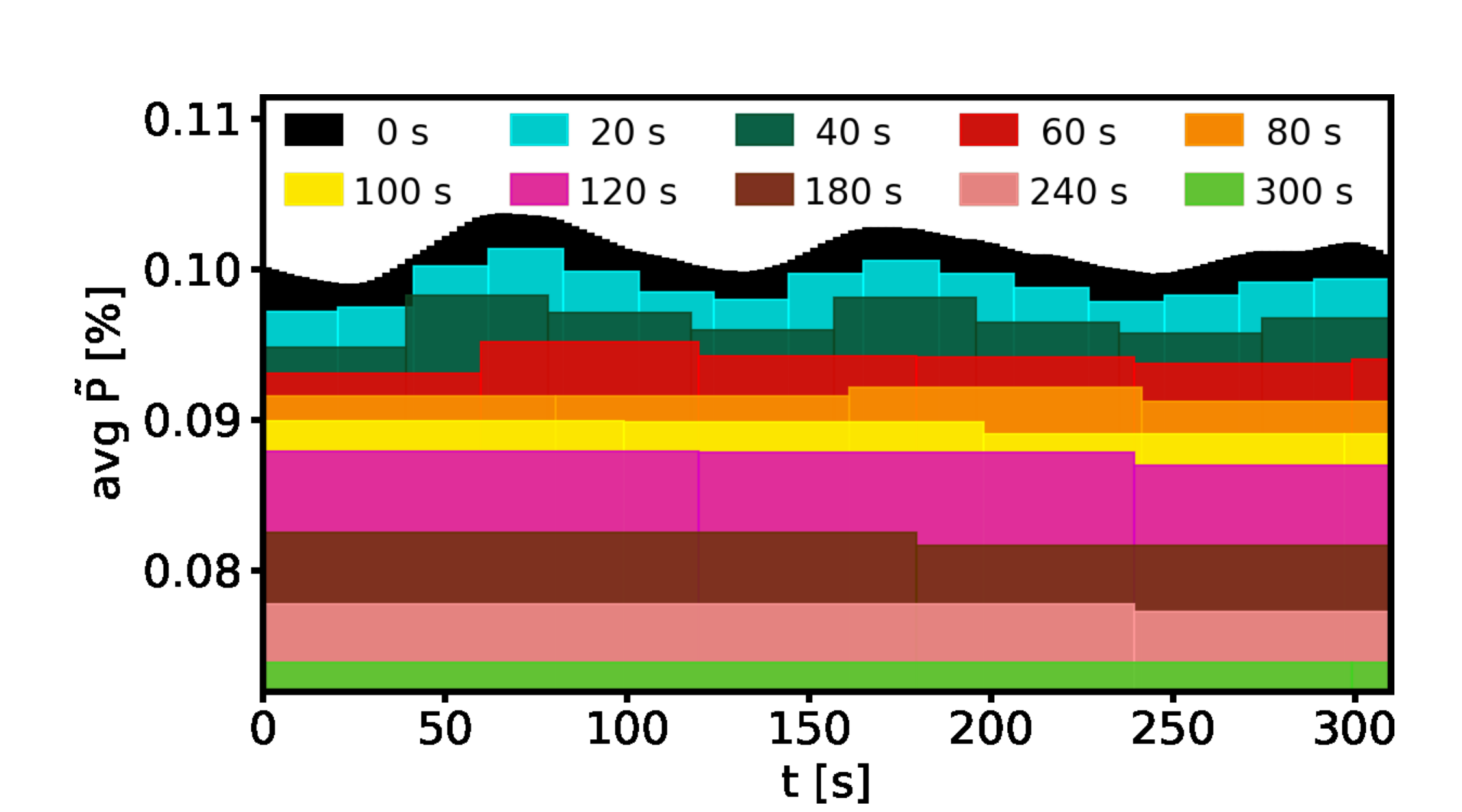}
\includegraphics[width=.45\textwidth]{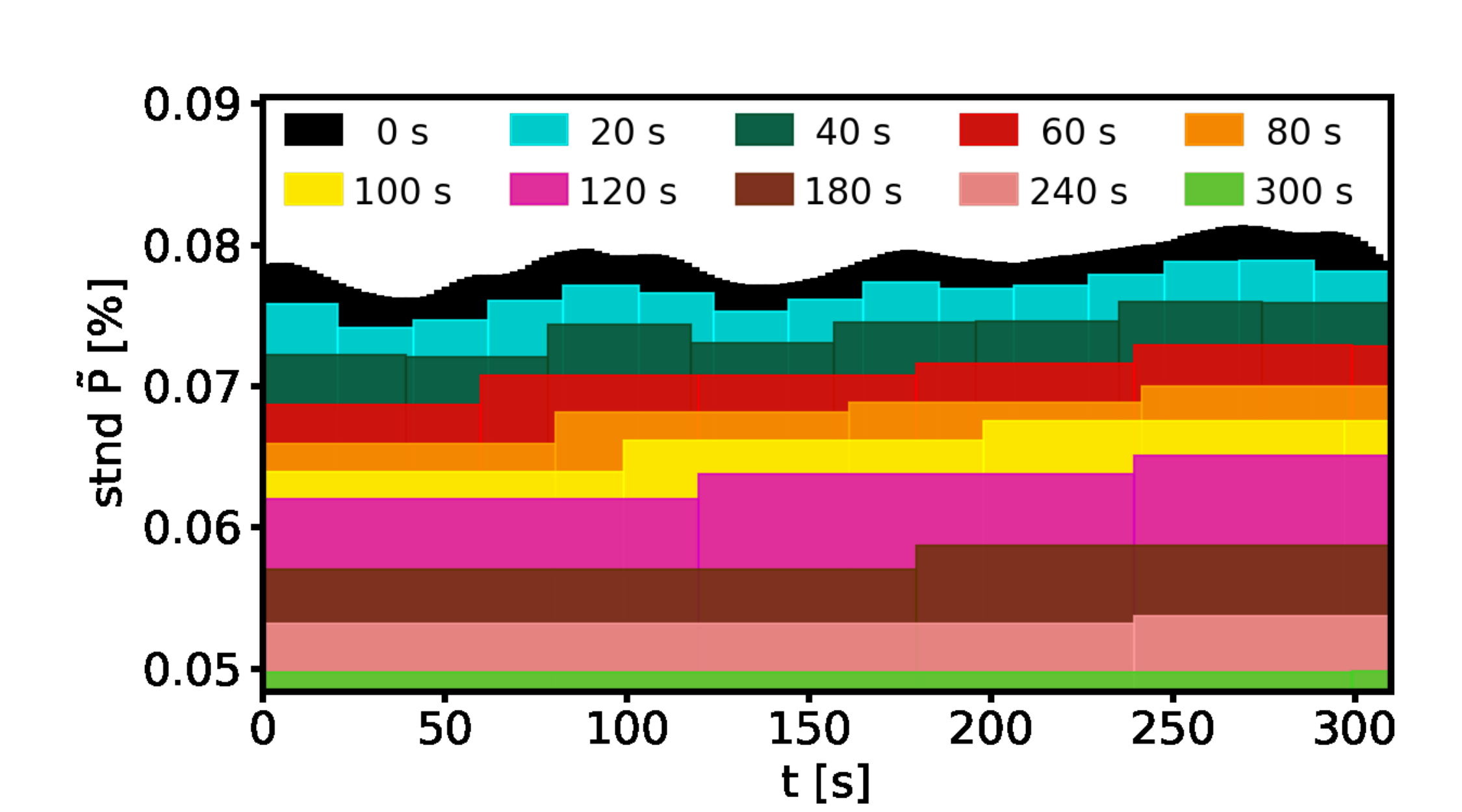}
\caption{Variation with time of the average value (left) and standard deviation
(right) of the total linear polarization at the line center, normalized to the
spatially and temporally averaged line-center intensity and for different
integration times (see legend).}
\label{F-timeintegralsstats} 
\end{figure}

Figure \ref{F-timeintegralsstats} shows the time evolution of the mean value
and the standard deviation of the total linear polarization in the time series
after integrating for different exposure times (see legend in the figure). The
chosen average $\tilde{P}$ is given by
\begin{equation}
\tilde{P} = \frac{\sqrt{Q^2 + U^2}}{\bar{I}} ,
\label{E-Pdef}
\end{equation}
with $\bar{I}$ the average intensity value, in time and space, of the whole
series, used for all time steps and integration times. All Stokes
parameters are taken at line center, for the disk-center line of sight. We have
chosen this polarization quantity because it is unsigned; that is, the mean
value of the individual $Q$ and $U$ signals is clearly affected by sign
cancellation, but $\tilde{P}$ is not. Additionally, the standard deviation
of $\tilde{P}$, $Q$, and $U$ show the same behavior.

Without taking into account any other instrumental effects besides the limited
time resolution, the average value and standard deviation of the total linear
polarization is already reduced $\sim10$~\% for one minute of exposure time;
that is, even a perfect instrument without any degradation of the spatial resolution
cannot fully detect the instantaneous polarization signals due to the finite
exposure time.

\section{Instrumental Effects}\label{Sobs}

In this section we study the effect on the intensity and linear polarization
of the \ion{Sr}{1} 4607~\AA\ line caused by the plasma evolution during the
finite exposure time (section \S\ref{Stime}), but including now the instrumental
effects (see \citealt{delPinoetal2018} for an analysis of instrumental effects
without the impact of time evolution for the same spectral line). We consider
two relevant instrumental setups: a filter polarimeter and a slit-based
spectropolarimeter.

\subsection{Filter polarimeter}  

This instrumental setup is of particular interest because it is similar to
that used by \cite{Zeuneretal2020}. Although they could not directly observe
the linear polarization pattern predicted by our disk-center simulations
(Fig. \ref{F-timeintegrals}), they were able to detect the expected linear
polarization signals after applying a novel analysis technique. We want to
test what happens to the time series simulation when we mimic the degradation
produced by a Fabry--P\'erot instrument similar to theirs. Our degradation
steps are the following:

\begin{itemize}
\item Telescope of 0.76~m diameter and $\sim0.5$" spatial resolution
($r_0 = 0.23$~m), computed with the long exposure MTF from \cite{Fried1966}.
\item Spatial sampling of 0.2"/pix (resulting from a 3x3 bining as in
\citealt{Zeuneretal2020})
\item Gaussian filter with 67~m\AA\ of full width at half maximum (FWHM).
As in \citet{Zeuneretal2020}, this filter is shifted 20~m\AA\
to the blue of the theoretical \ion{Sr}{1} 4607~\AA\ line center.
\item Noise in Stokes $Q$ and $U$ from a normal distribution with
$\sigma = 4\cdot10^{-3}I$ (hereafter, 0.4~\%) for each snapshot
($\sim 2$~s exposure), which corresponds to approximately $0.04$~\% for a
$3.5$~min integration.
\end{itemize}

\begin{figure}[htp]
\centering 
\includegraphics[width=.95\textwidth]{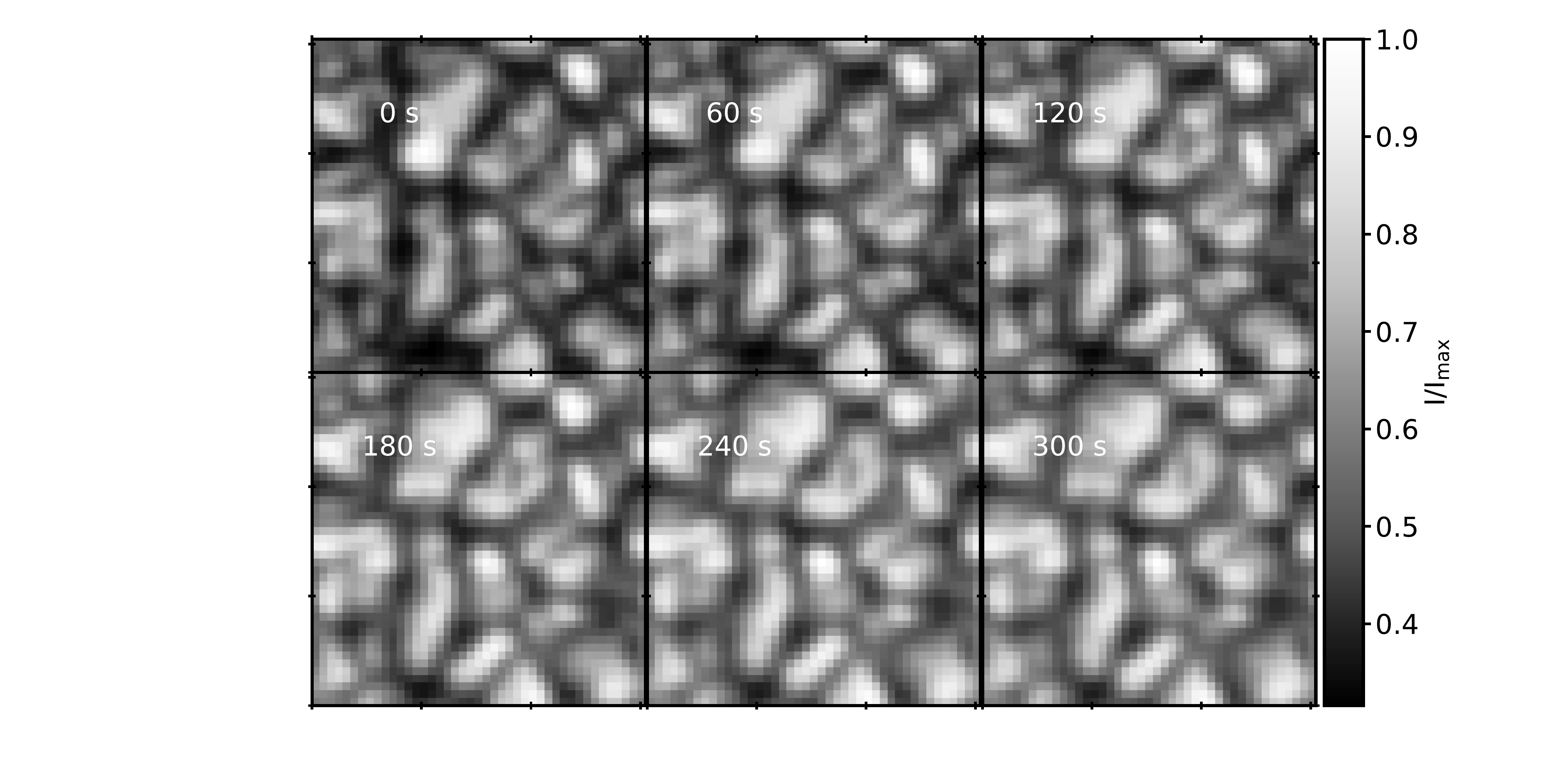} \\\vspace*{-2.85em}
\includegraphics[width=.95\textwidth]{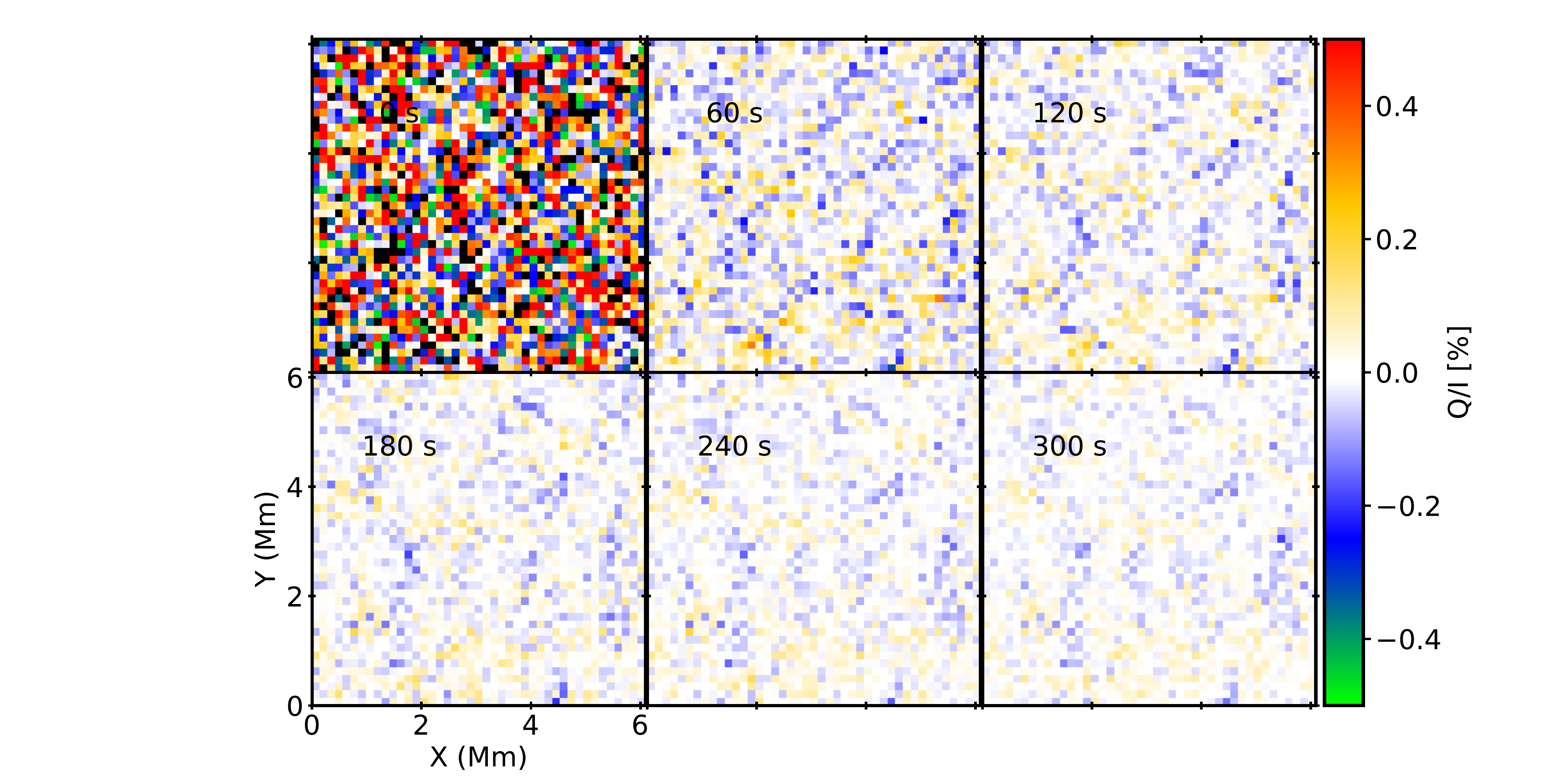}
\caption{
Intensity (top panels) and fractional linear polarization $Q/I$
(bottom panels) at the \ion{Sr}{1} 4607~\AA\ line center, for the disk-center LOS, for
the first snapshot of the time series and for different integration times,
degraded as if observed by the filter polarimeter described in the text, from
top to bottom and left to right: $0\,$s, $60\,$s, $120\,$s, $180\,$s, $240\,$s,
and $300\,$s.}
\label{F-timeintegrals-filter} 
\end{figure}

Once we degraded all the snapshots following these steps, we proceeded to
integrate them in time to simulate different exposure times.
Figure \ref{F-timeintegrals-filter} shows the intensity and fractional linear
polarization $Q/I$ for the filter polarimeter described above. The
intensity (top panels) has been normalized to the maximum for each panel. We
note that the increase of integration time decreases the contrast
($(I_{\rm max}-I_{\rm min})/(I_{\rm max}+I_{\rm min})$, from $0.52$ to $0.45$
between the first and last panel). For the polarization we have limited the
plotting range of values to $\left[-0.5,0.5\right]$. This improves the
visualization of most of the panels, but makes it so that the first panel is
saturated. However, the first panel shows pure noise. We can see that a five
minutes integration returns a pattern that reminds us of the original snapshot
(see fig. \ref{F-timeintegrals}), but the amplitude of the signal is
decreased significantly.

\begin{figure}[htp]
\centering 
\includegraphics[width=.95\textwidth]{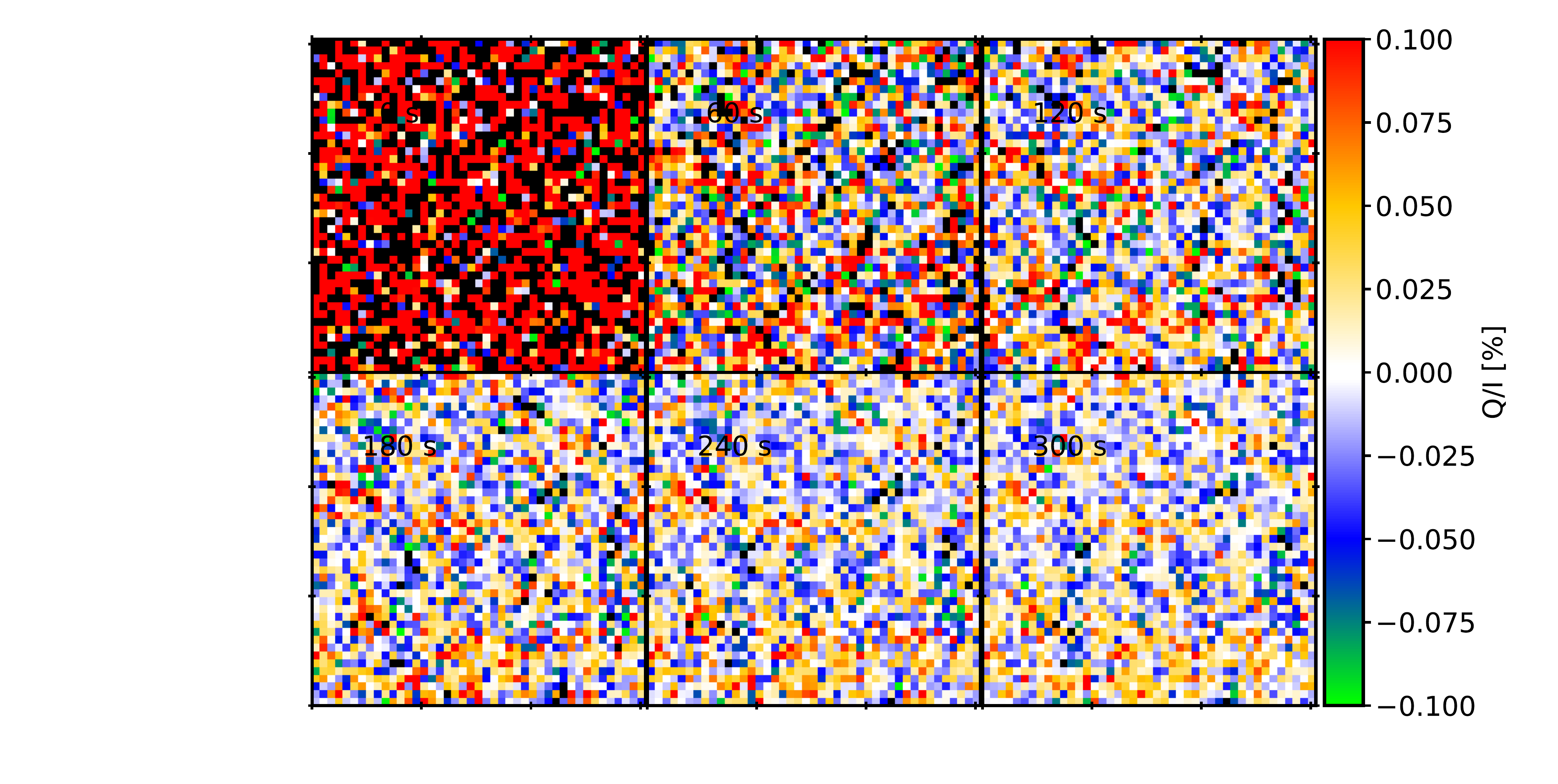}
\includegraphics[width=.95\textwidth]{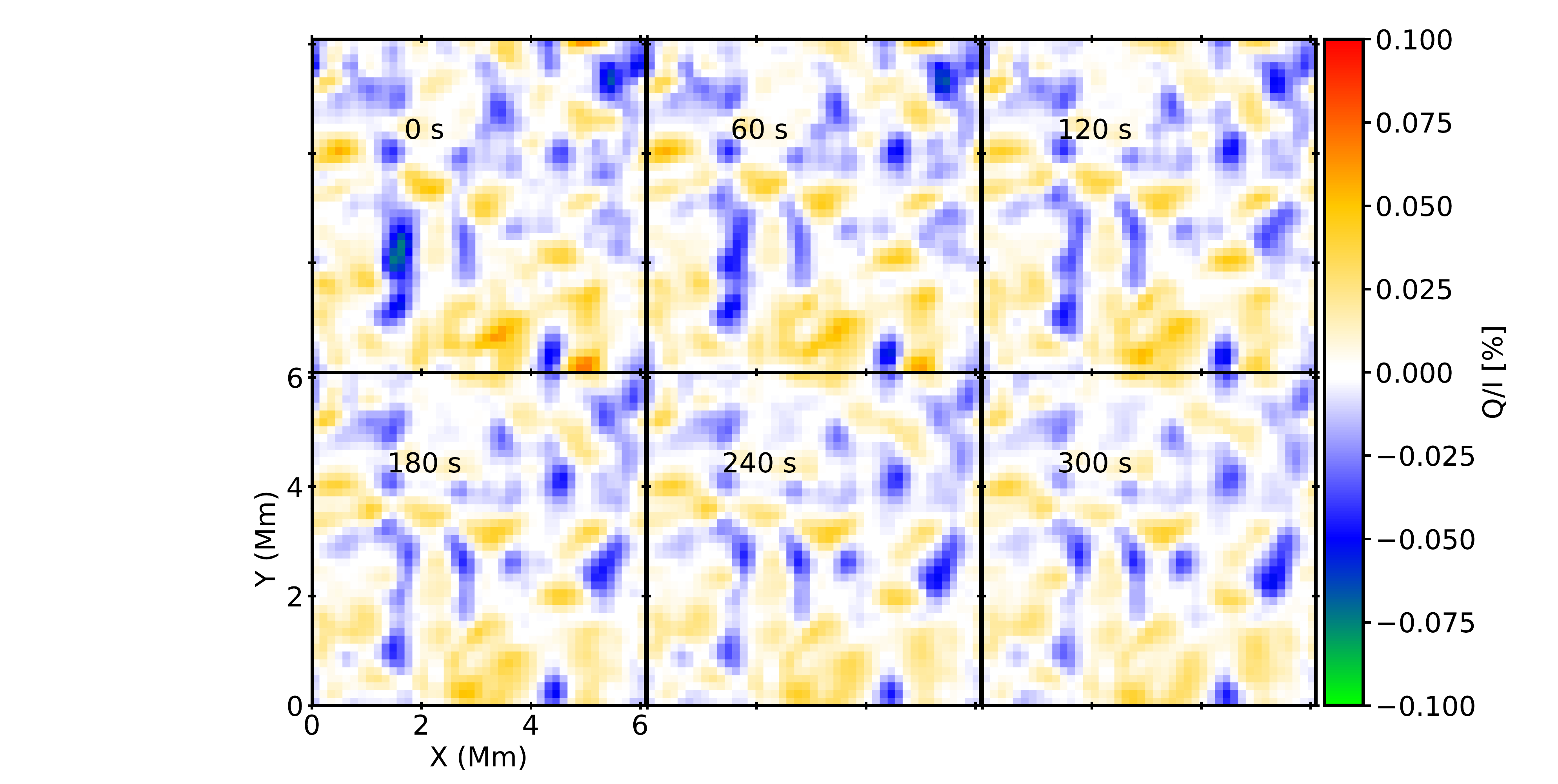}
\caption{The top panels show the same $Q/I$ panels of Fig.
\ref{F-timeintegrals-filter} but changing the color scale for an easier comparison
with the noise-free case shown in the bottom panels.}
\label{F-timeintegrals-filter-nonoise} 
\end{figure}

However, a closer look into the details in the bottom panel of Fig.
\ref{F-timeintegrals-filter} shows that, even for a $5$~min integration,
the resulting linear polarization is severely affected by the noise (see also the
top panels in fig. \ref{F-timeintegrals-filter-nonoise}, where we have saturated
the color scale for an easier visualization).  It is also of interest to show the
disk-center fractional linear polarization map that the filter polarimeter would
observe if there were no noise (see bottom panel in Fig.
\ref{F-timeintegrals-filter-nonoise}). A comparison with the bottom panel in Fig.
\ref{F-timeintegrals} shows that in the noise-free ideal case the instrumental setup
used by \cite{Zeuneretal2020} would reduce the original theoretical polarization
amplitudes by more than an order of magnitude, but would keep the structure of the
predicted polarization pattern. However, as seen in the top panel of Fig.
\ref{F-timeintegrals-filter-nonoise}, the noise level is such that this pattern is
almost lost even for a five minutes integration time.

\begin{figure}[htp]
\centering 
\includegraphics[width=.45\textwidth]{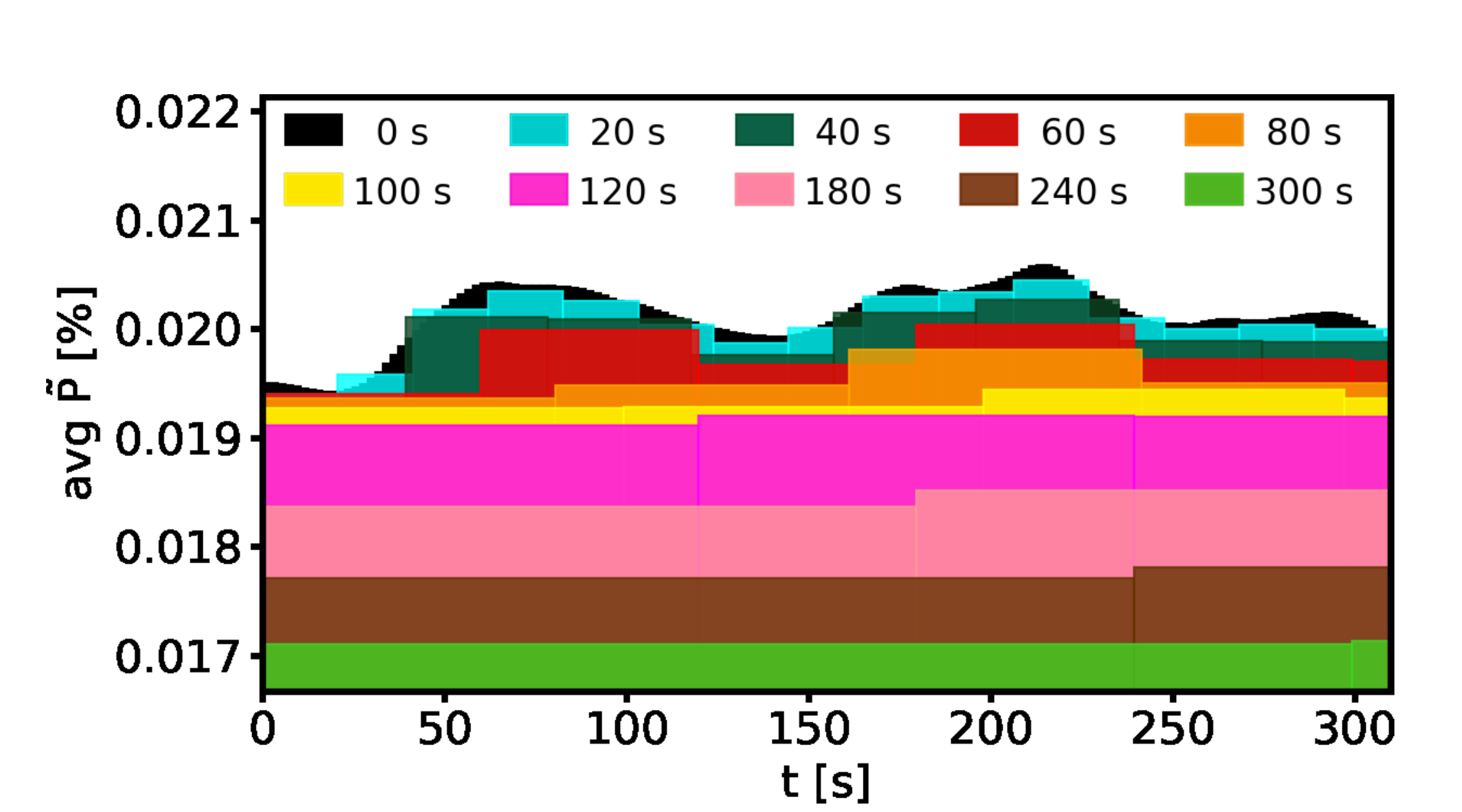}
\includegraphics[width=.45\textwidth]{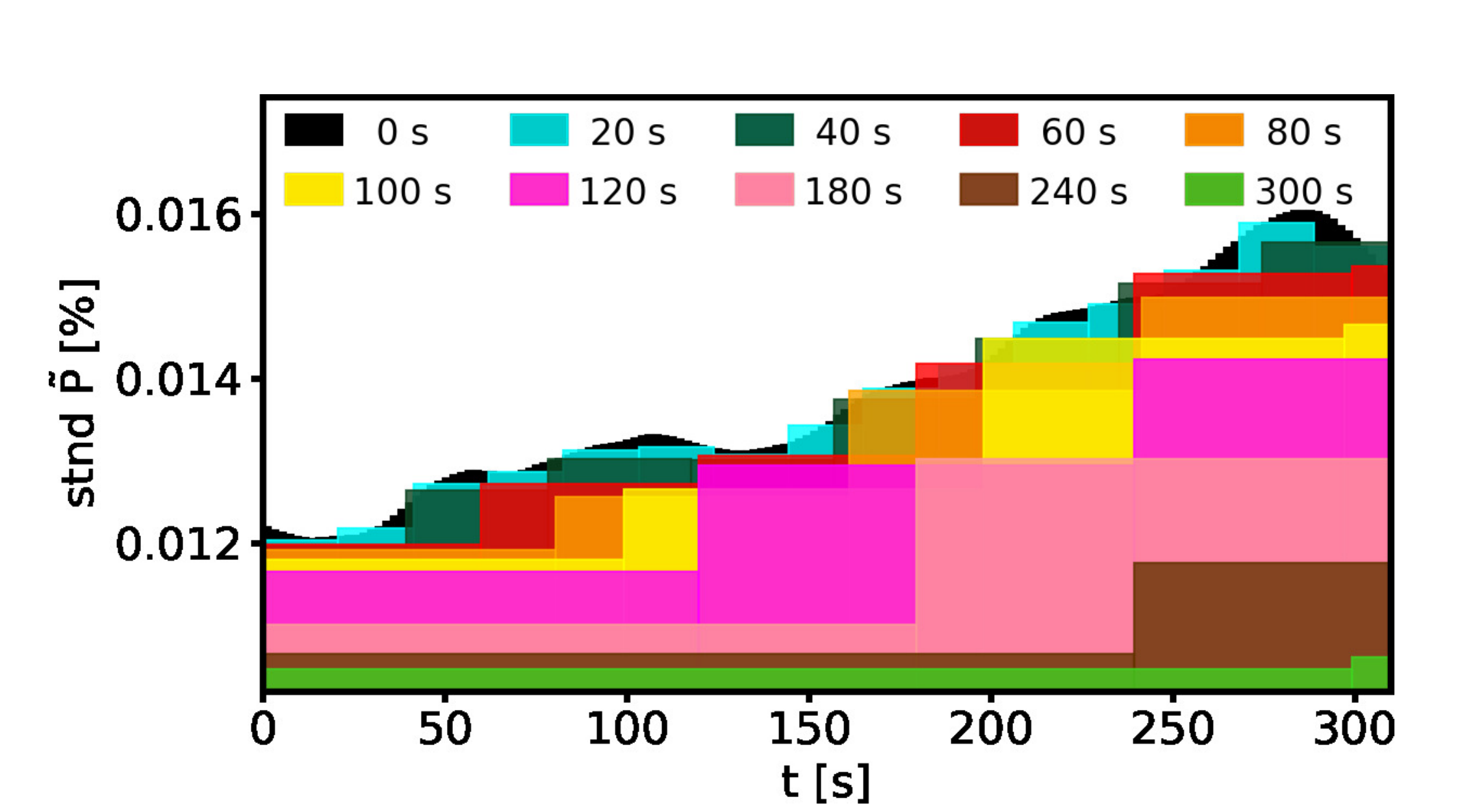} \\
\includegraphics[width=.45\textwidth]{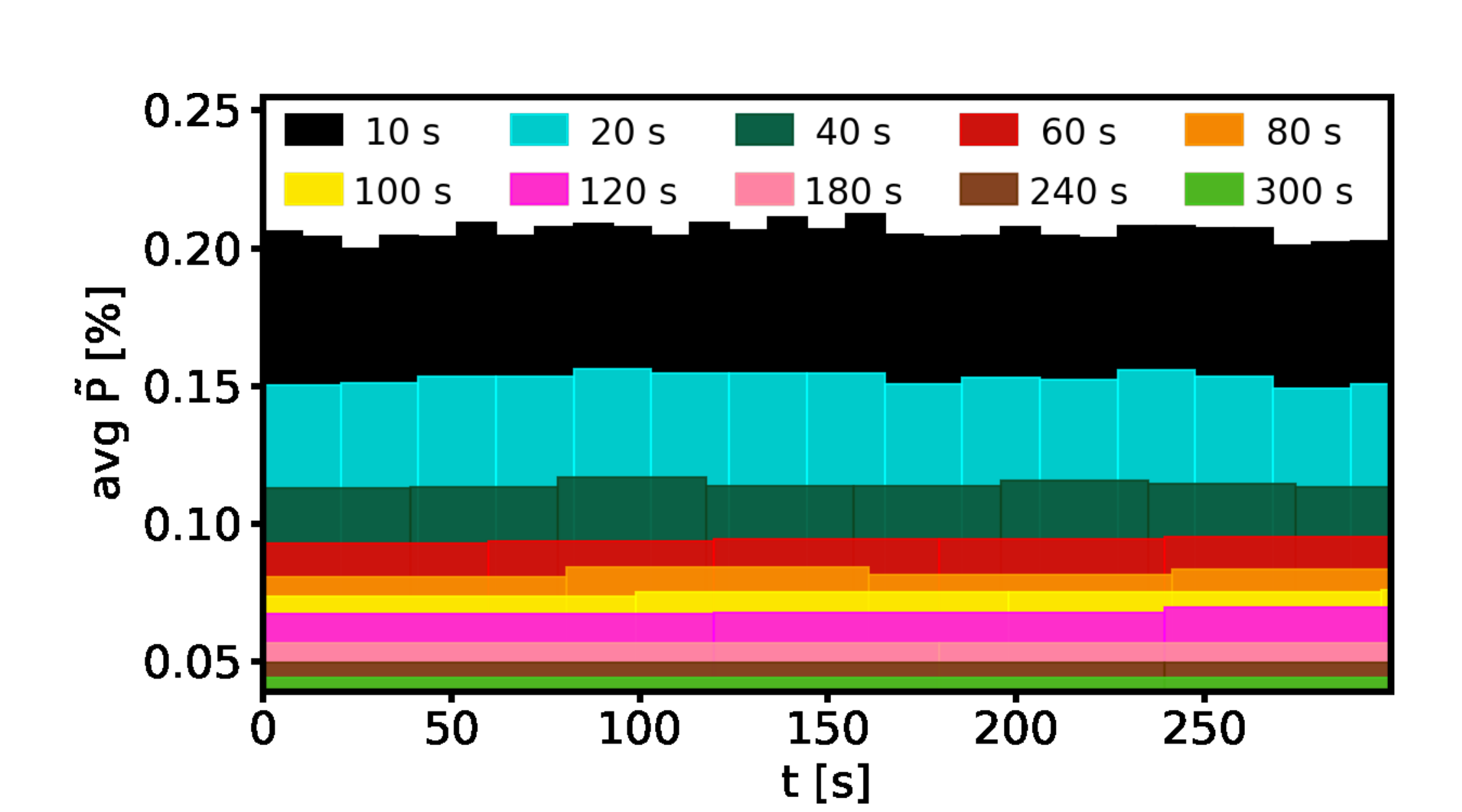}
\includegraphics[width=.45\textwidth]{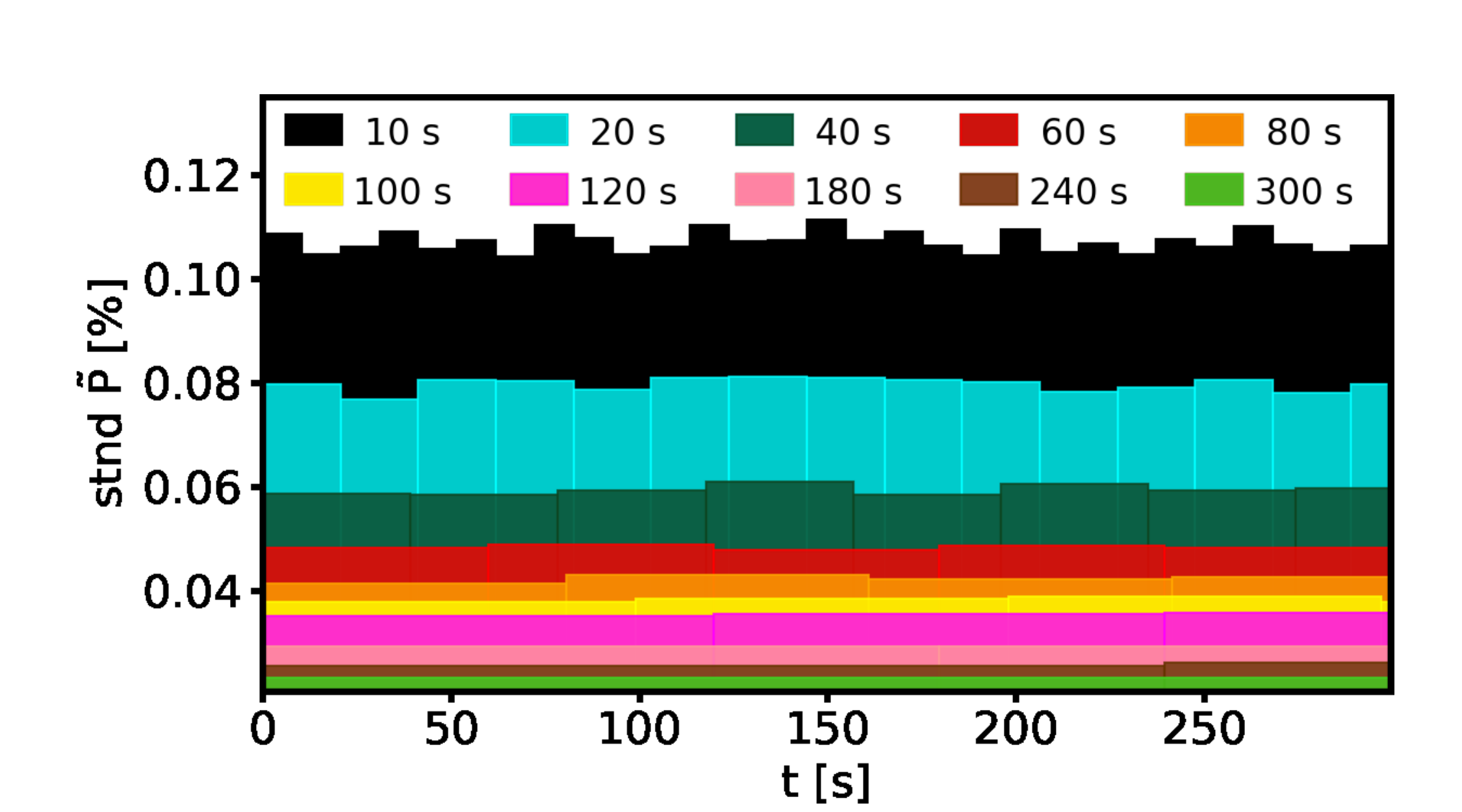}
\caption{
Variation with time of the average value (left) and standard deviation
(right) of the total linear polarization, normalized to the spatially and
temporally averaged intensity, for different integration times (see legend)
for the filter polarimeter described in the text. Top panels: without noise.
Bottom panels: with noise.}
\label{F-timeintegralsstats-filter} 
\end{figure}

Regarding the time evolution of the mean value and the standard deviation
of the total linear polarization in the degraded time series, for different exposure
times (see Fig. \ref{F-timeintegralsstats-filter}) we see that the effect
of the noise is critical for the statistics of the observation. In the noise-free
case (top panels), we see some time evolution (notice, however, that the
range of variation in the figure is very small, namely, $\sim$0.004~\%) and mean
and standard deviation values much smaller than in the original simulation
(fig. \ref{F-timeintegralsstats}). This decrease is a consequence of the
simulated instrumental effects. However, when noise is included it dominates
the time evolution and the statistics, especially for small integration times.
For this reason, both the mean and standard deviation quickly decrease with
the integration time when the noise is taken into consideration
(see the bottom panels of Fig. \ref{F-timeintegralsstats-filter}).

\begin{figure}[htp]
\centering 
\includegraphics[width=.65\textwidth]{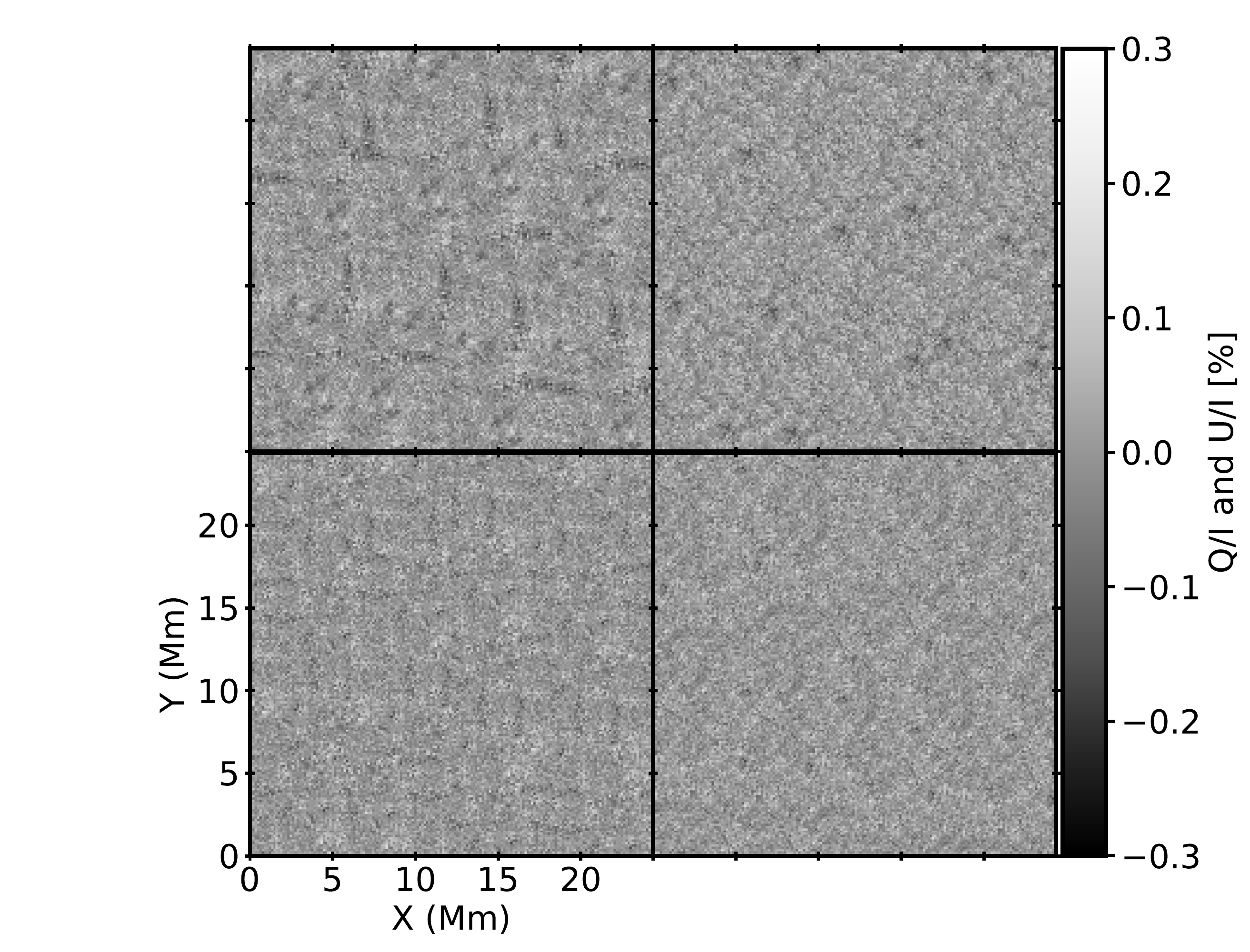}
\caption{
Fractional linear polarization $Q/I$ (left column) and $U/I$ (right column)
for the first snapshot in the series, without time evolution but with a noise value
corresponding to a $210$~s exposure (top row) and for an exposure time of $210$~s
(bottom row), for the filter polarimeter described in the text,
with the same scale as Fig.~$1$ in \cite{Zeuneretal2020}.}
\label{F-timeintegrals-filter-compare} 
\end{figure}

In order to compare with the observations by \cite{Zeuneretal2020} we show
the case of $210$~s exposure time in the same amplitude and color scale
used by them (see bottom panels in Fig. \ref{F-timeintegrals-filter-compare}).
To study the effect of the time evolution on the polarization amplitudes,
we have taken the initial snapshot applying to it the same degradation
procedure, but with a polarization noise with $\sigma = 0.4$~\%
(the equivalent value for a $210$~s integration). In order to present a
field of view similar to the one of their observation, we have replicated
16 times the field of view of the model's snapshots; however, we have
introduced some shifts between the pieces of the mosaic to avoid the
obvious pattern repetition that we would have if we just replicated the
same plane. It seems, from visual comparison with \cite{Zeuneretal2020},
that we get slightly larger polarization amplitudes than in their
$U/I$ observation, consistent with their finding regarding the
noise distribution difference between $Q/I$ and $U/I$ explained in
appendix B of \cite{Zeuneretal2020}. Moreover, although there are some
differences between the top and bottom panels in Fig.
\ref{F-timeintegrals-filter-compare}, we can see that taking into
account the exposure time of their $210$~s observation does not
dramatically impact the average amplitude (as suggested by the noise-free
average in the top left panel of Fig. \ref{F-timeintegralsstats-filter}).
However, the amplitude is dominated by noise (see
Fig. \ref{F-timeintegrals-filter-nonoise}). Note that, even though
we have assumed a duty cycle of 100~\% when integrating the different
snapshots, the comparison with \cite{Zeuneretal2020} is still valid
because the FSP 2 has a duty cycle of 98~\% \citep{Iglesiasetal2016}.

\subsection{Slit-based spectropolarimeter}

Our second instrumental setup is the Visible Spectro-Polarimeter (ViSP)
at the DKIST (e.g., \citealt{Elmoreetal2014}). We have chosen this slit-based
spectropolarimeter because it is of present interest to predict what type of
scattering polarization signals we can expect to observe using the ViSP at
the world's largest solar telescope. Our degradation steps are the following:

\begin{itemize}
\item Telescope of 4~m diameter and $\sim0.1$" spatial resolution
($r_0 = 1.16$~m) computed with the long exposure MTF from \cite{Fried1966}.
Clearly, such spatial resolution requires the use of adaptive optics.
\item Spectral point spread function, convolution with a gaussian with
33.69~m\AA\ FWHM.
\item Spectrograph slit with a $0.1071$" width (this results in $\sim0.2$"
spatial resolution in the scanning direction).
\item Spectral sampling of 34~m\AA\ /pixel and spatial sampling along the slit
of $0.06066$"/pixel.
\item Noise in Stokes $Q$ and $U$ from a normal distribution with
$\sigma = 0.04$~\% (relative to the continuum intensity) for each snapshot
($\sim 2$~s exposure), which corresponds to a signal to noise ratio (SNR) of
$\sim 2300$.
\end{itemize}

The numerical values for the ViSP degradation have been obtained using the
ViSP IPC\footnote{https://www.nso.edu/telescopes/dkist/instruments/visp/},
a calculator for observation planning.

Once we degraded all the snapshots following these steps, we
proceeded to integrate them in time to simulate different exposure times. In the
present ViSP slit spectropolarimeter case we have simulated the scan, i.e., the
exposure time corresponds to the time each slit is kept in position. Because
scanning a field of view of this size ($\sim 10$") with this slit width and
step (we move the slit a distance equal to its width, $\sim0.1$") requires more
than five minutes for reasonable exposure times, we quickly run out of snapshots to
integrate. To solve this, whenever we reach the end of the series, we
switch the direction of the time evolution, that is, the time series mirrors
itself whenever it reaches the last snapshot. Note, however, that we do not
take into account other effects such as the possible misalignments and
the finite time in the scan that should be devoted to the reading of the
cameras or the movement of the slit.

\begin{figure}[htp]
\centering 
\includegraphics[width=.95\textwidth]{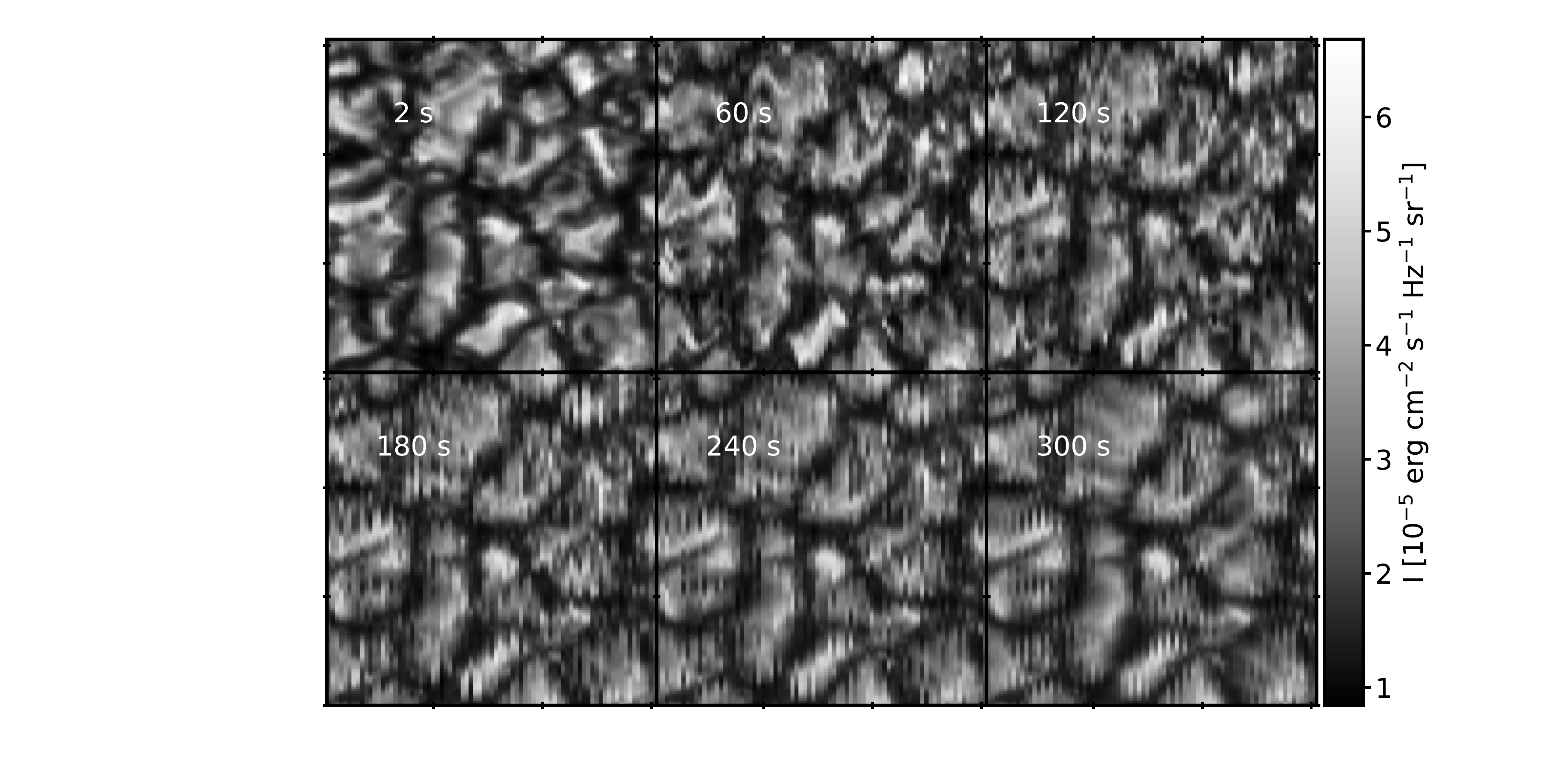} \\\vspace*{-2.85em}
\includegraphics[width=.95\textwidth]{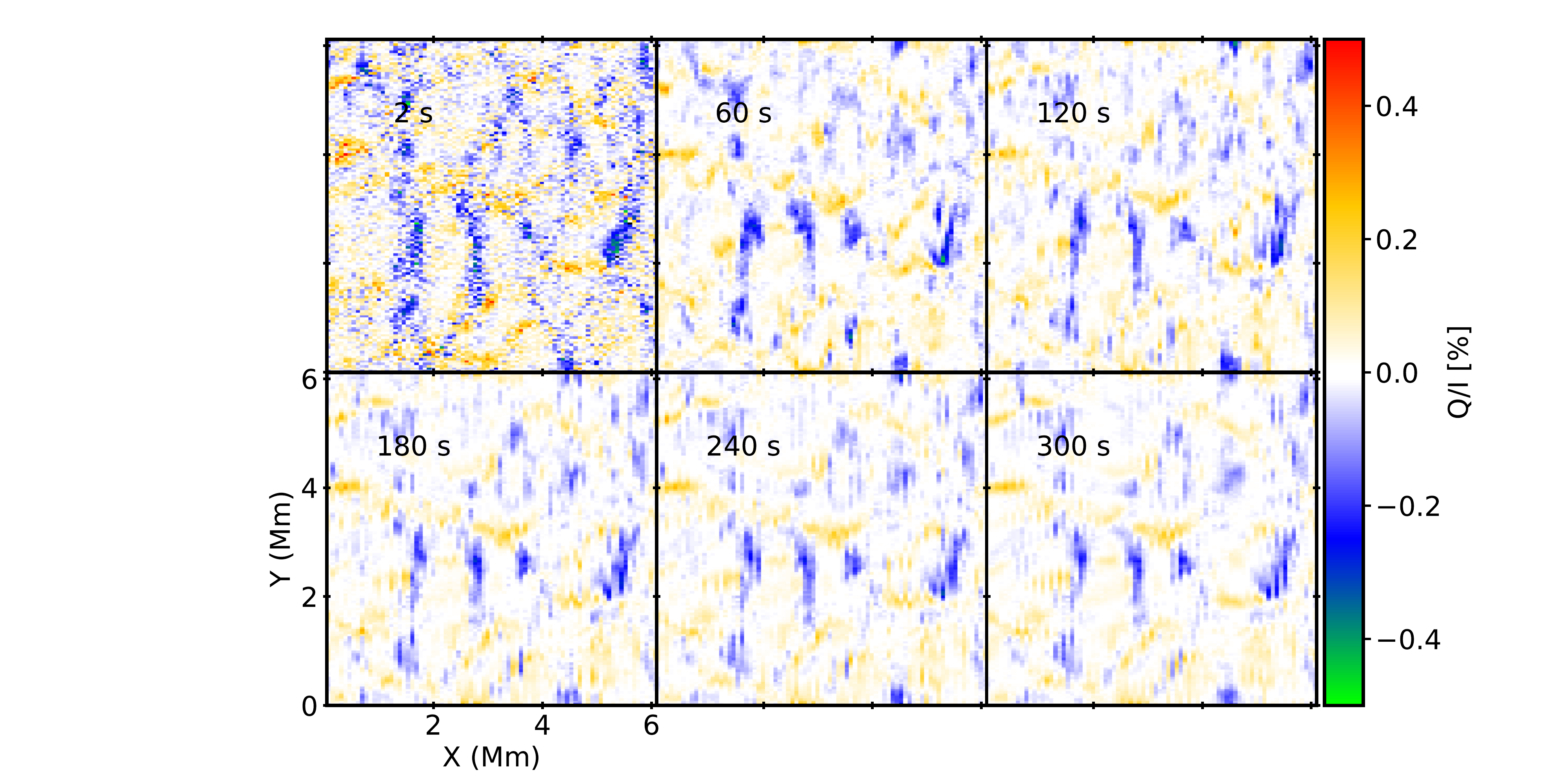}
\caption{
Intensity (top panels) and fractional linear polarization $Q/I$
(bottom panels) at the \ion{Sr}{1} 4607~\AA\ line center, for the disk-center LOS, for
the first snapshot of the time series and for different integration times,
degraded as if observed by the slit spectropolarimeter described in the text.
From top to bottom and left to right: $0\,$s, $60\,$s, $120\,$s, $180\,$s,
$240\,$s, and $300\,$s. The noise changes as
$\sim0.04\cdot\sqrt{\tfrac{2}{t [s]}}$~\% and, therefore, it goes from $0.04$~\% for the $2$~s
exposure to $\sim0.003$~\% for the $300$~s exposure.}
\label{F-timeintegrals-slit} 
\end{figure}

Figure \ref{F-timeintegrals-slit} shows the intensity and fractional linear
polarization $Q/I$ for the slit spectropolarimeter described above. As it
happened with the filter polarimeter, larger integration times result
in a decreased intensity contrast (from $0.77$ to $0.70$ between the first
and last Stokes $I$ panel). We can also see the effect of the scanning in the
reconstructed image when we increase the integration time. This effect is, in fact,
underestimated, as our time series is roughly five minutes long and we have
repeated it as many times as necessary to cover the whole simulation field
of view with the chosen slit. If we had a long enough series, the difference
in the granulation pattern between the reconstructed intensity and the result
of the simulation would be more significant. On the other hand, the much
higher SNR for the $4$~m telescope is very apparent, as only in the
instantaneous snapshots can we clearly see the noise, greatly diminished already
after a $30$~s integration. Notice that even with this relatively high spatial
resolution ($0.1$"$\times0.2$") the simulated signal is roughly a factor $2$
smaller than the one obtained in the simulation.

\begin{figure}[htp]
\centering 
\includegraphics[width=.45\textwidth]{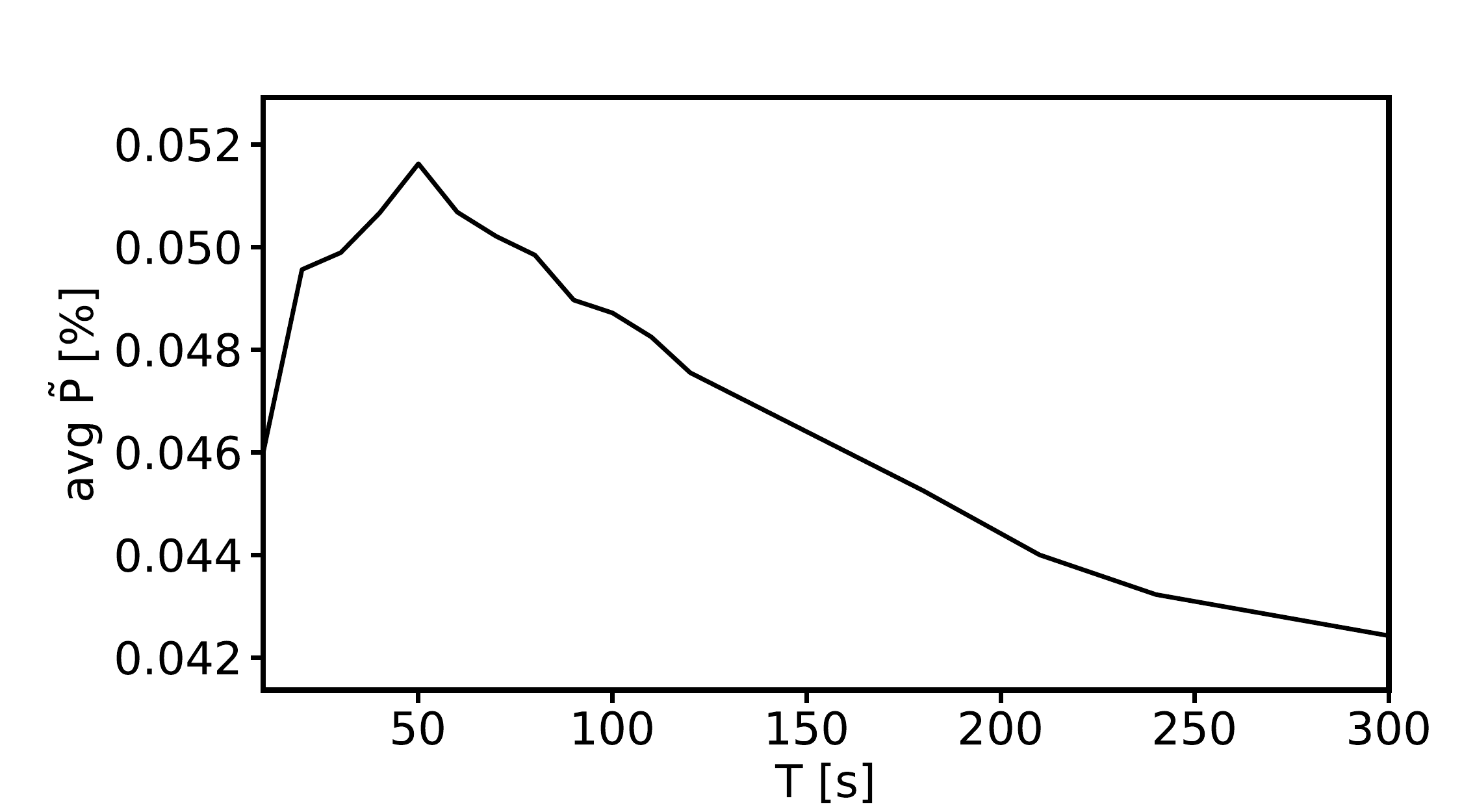}
\includegraphics[width=.45\textwidth]{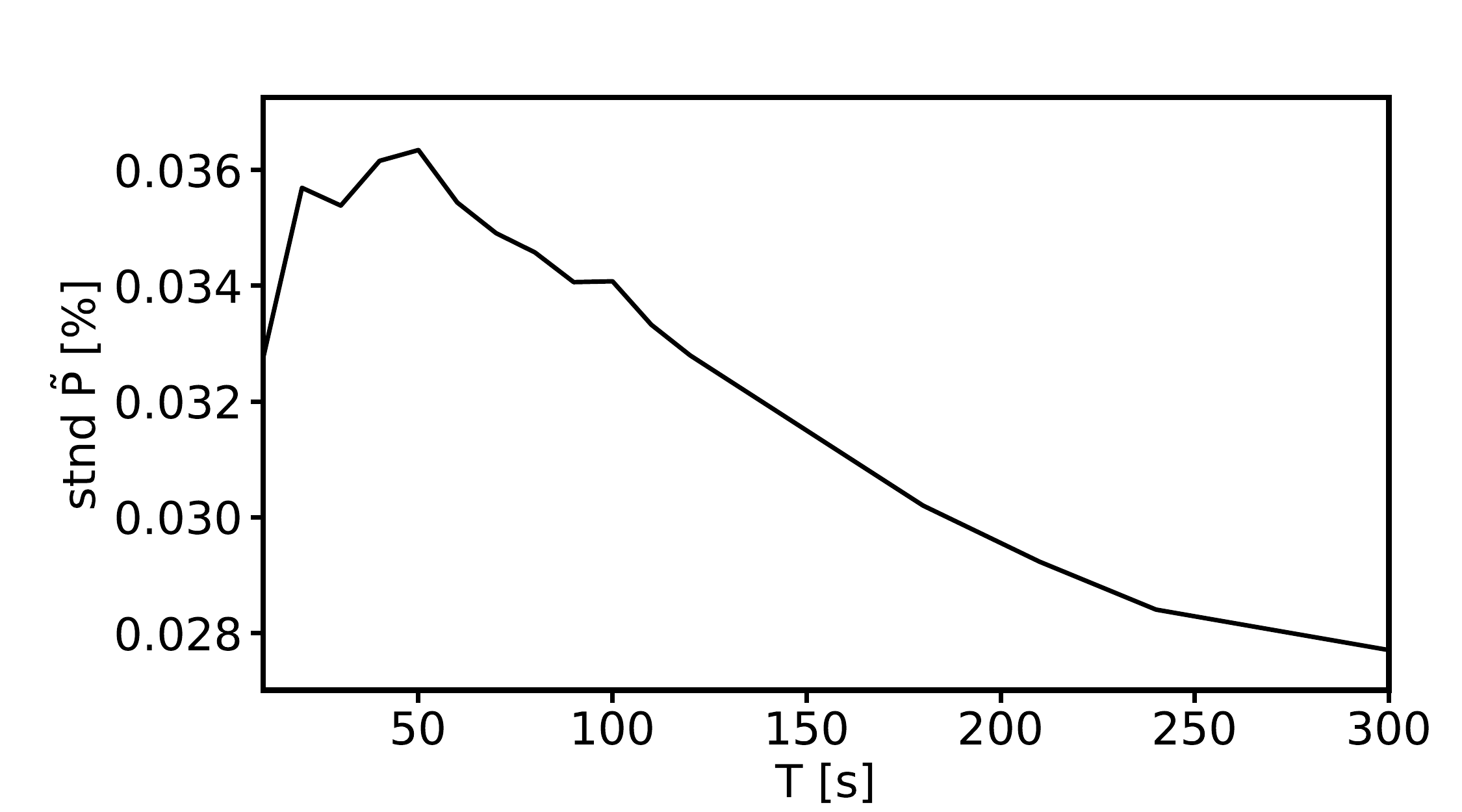} \\
\includegraphics[width=.45\textwidth]{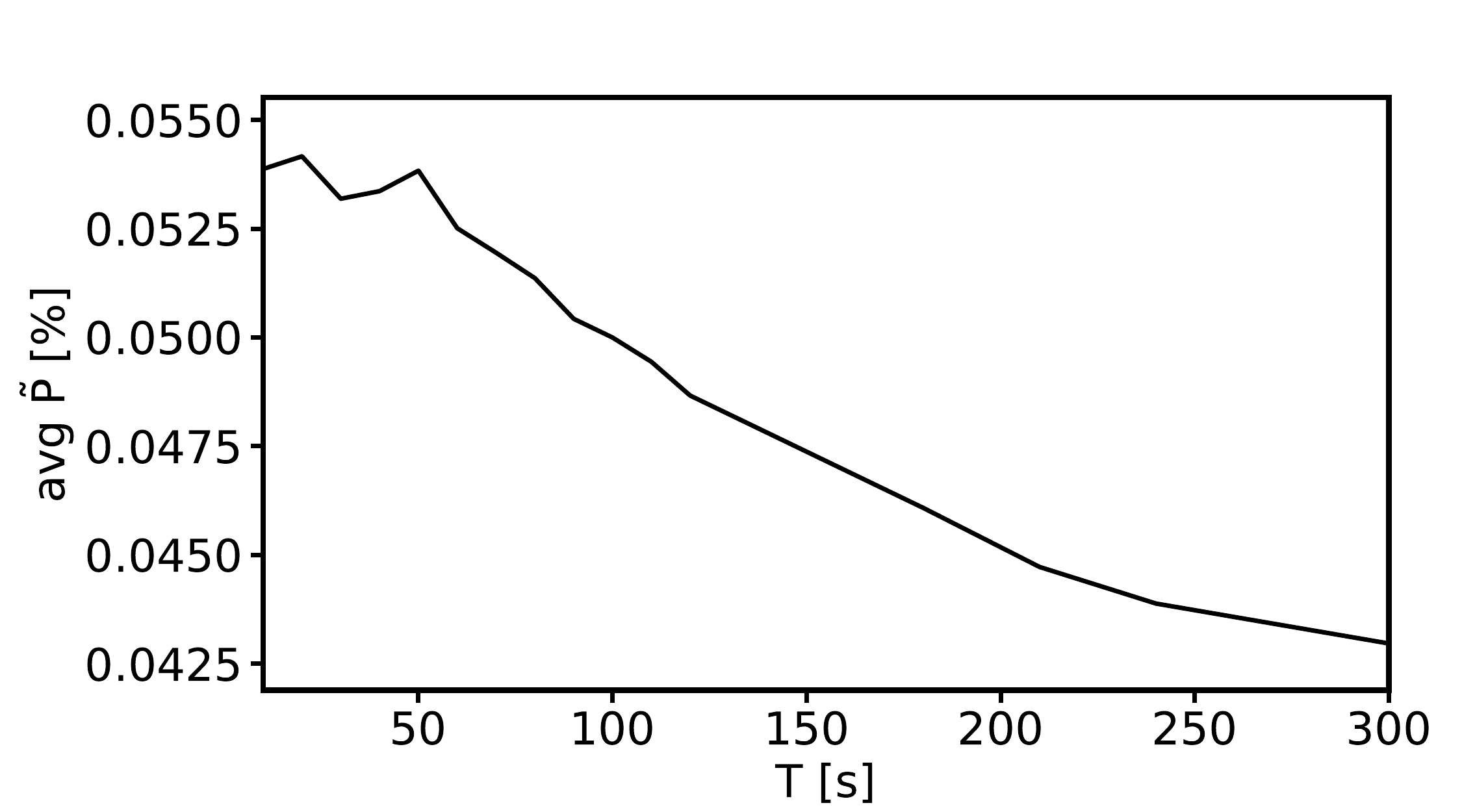}
\includegraphics[width=.45\textwidth]{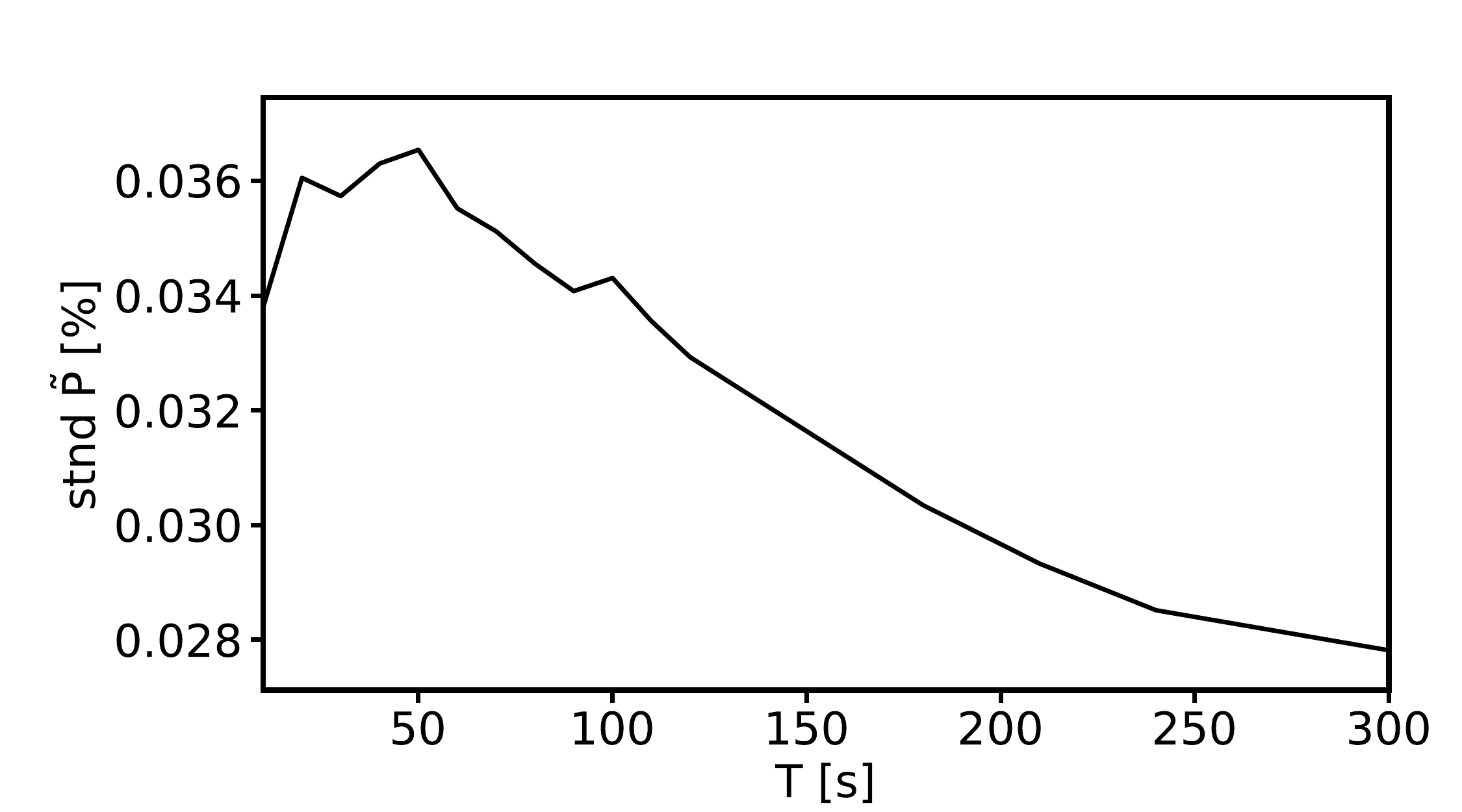}
\caption{
Variation with the integration time of the average value (left) and
standard deviation (right) of the total linear polarization, normalized to
the spatially and temporally averaged intensity,
for the slit spectropolarimeter described in the text.
Top panels: without noise.  Bottom panels: with noise.}
\label{F-timeintegralsstats-slit} 
\end{figure}

Figure \ref{F-timeintegralsstats-slit} shows the variation of the mean
value and the standard deviation of $\tilde{P}$ (Eq. \eqref{E-Pdef}) for
different integration times. Because with this instrumental setup we need
to scan along one of the spatial directions, there is no time evolution.
The initial rise on the statistics can be due to their temporal evolution,
as both the mean and standard deviation increase during the first $60$~s
in the synthetic maps (see Fig. \ref{F-timeintegralsstats}). After this
initial increase, the curves are dominated by the time integration and both
the average and standard deviation values decay resembling an order three
polynomial. Because this setup simulates a $4$~m telescope working at $0.1$"
resolution, the SNR is much higher and the impact of the noise is much
smaller than with the filter polarimeter setup (compare top and
bottoms panels in Fig. \ref{F-timeintegralsstats-slit}).

\begin{figure}[htp]
\centering 
\includegraphics[width=.45\textwidth]{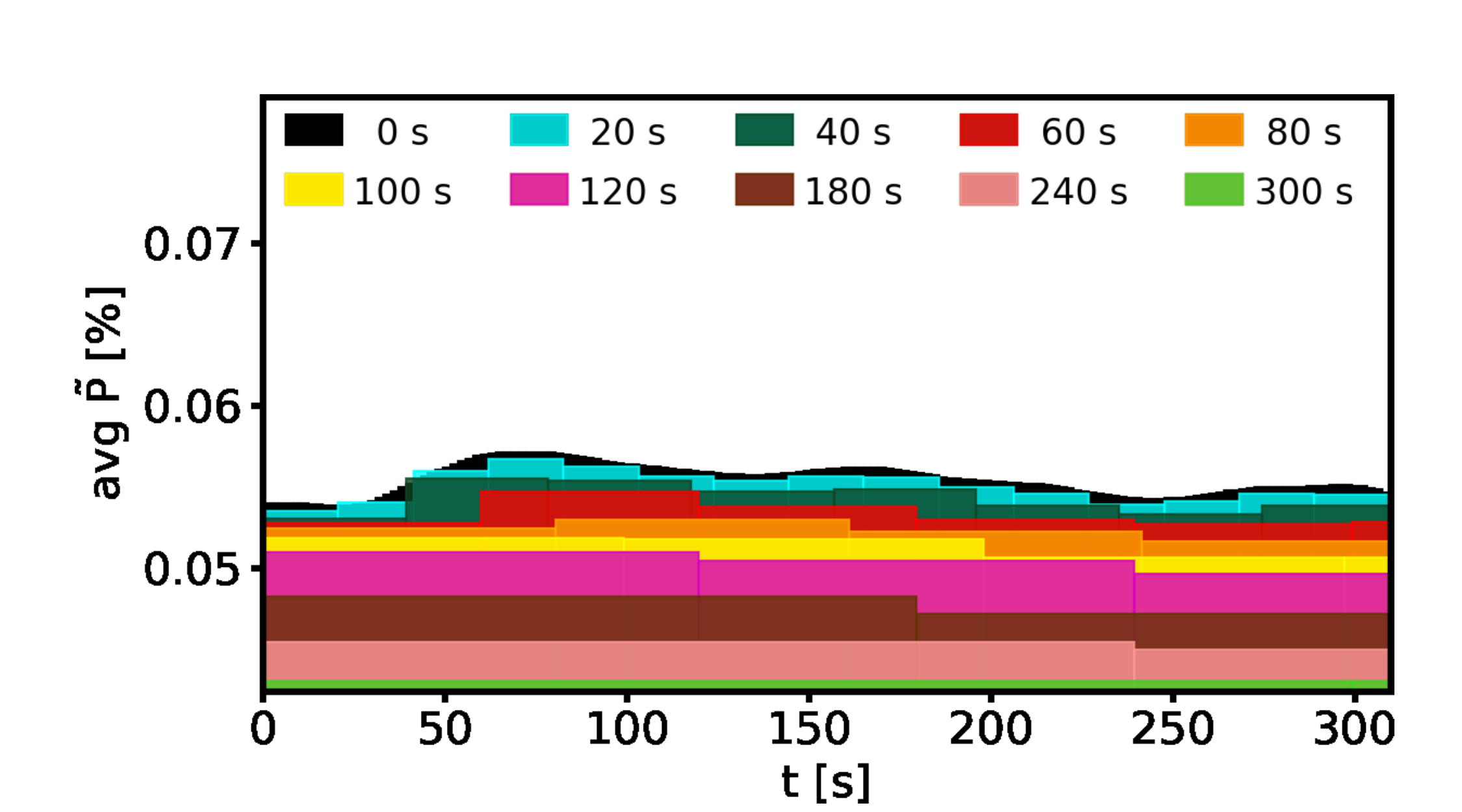}
\includegraphics[width=.45\textwidth]{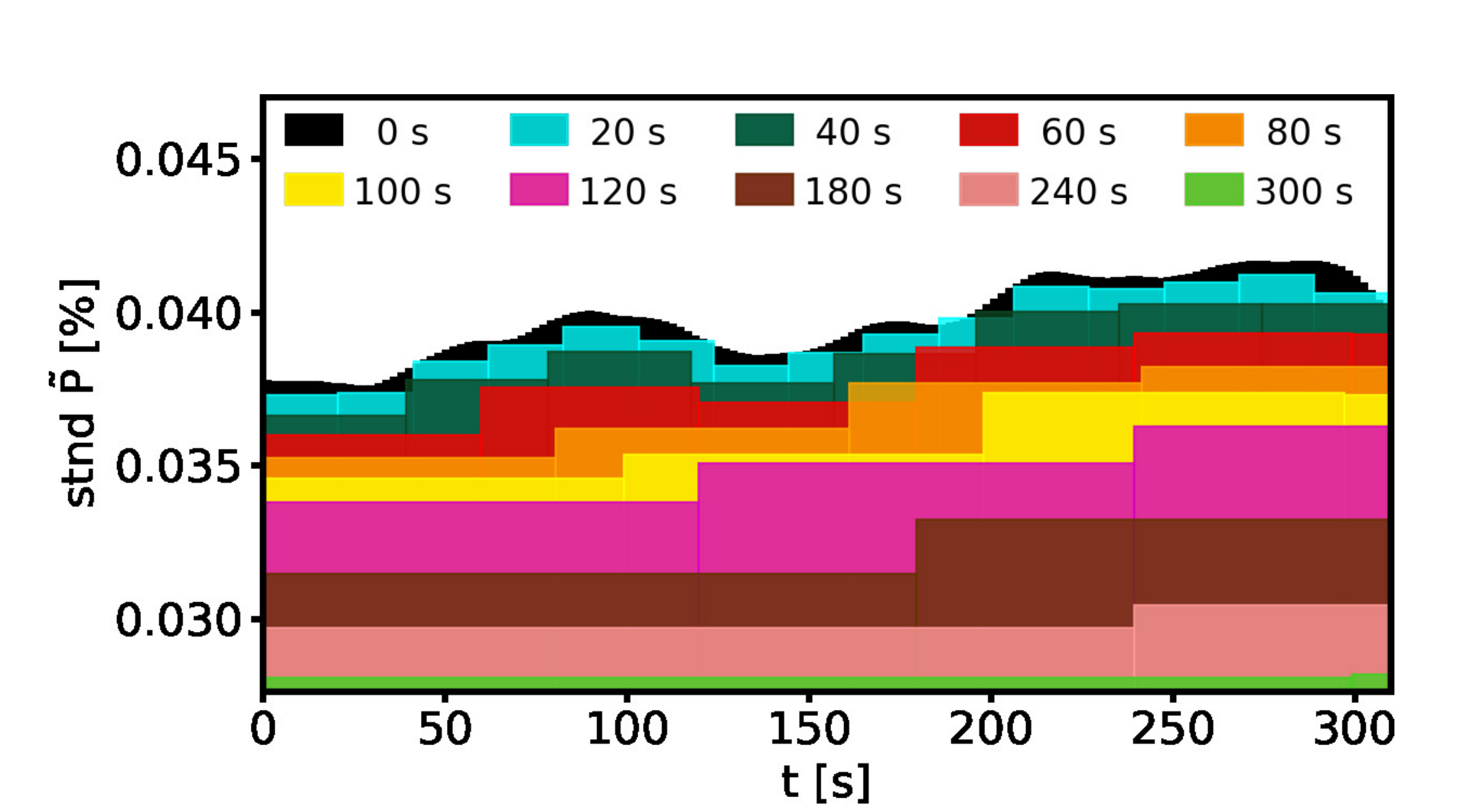} \\
\includegraphics[width=.45\textwidth]{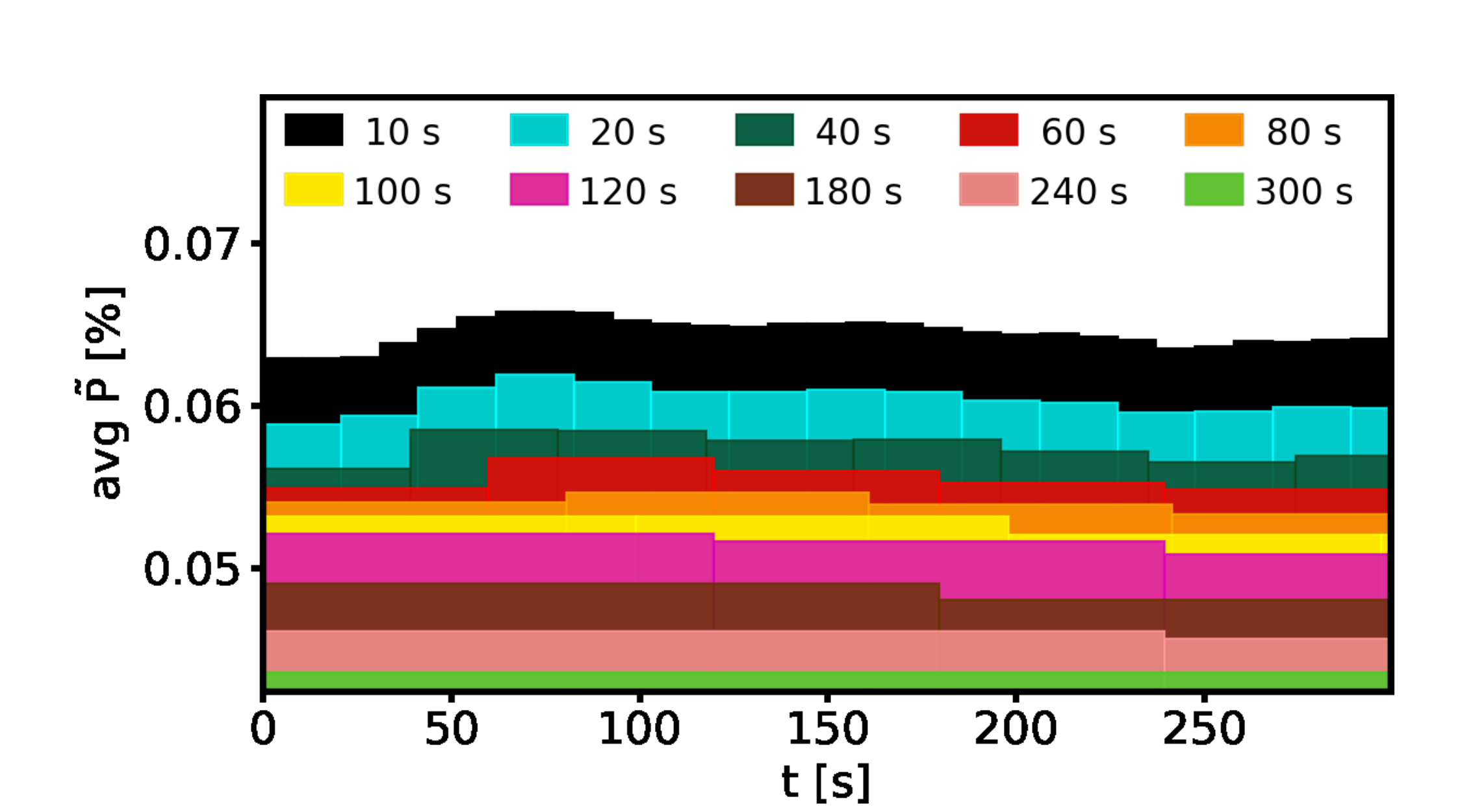}
\includegraphics[width=.45\textwidth]{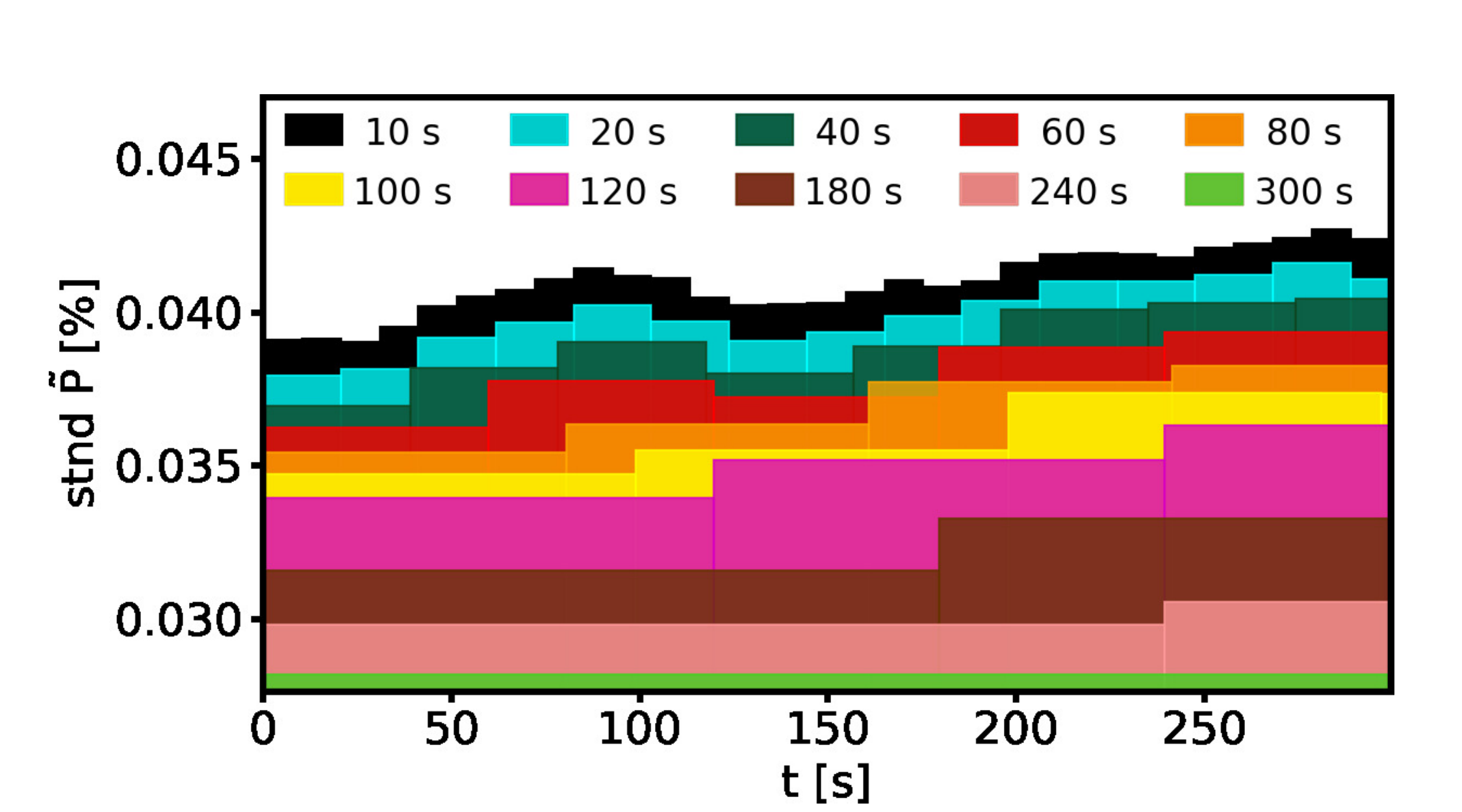}
\caption{
Variation with time of the average value (left) and standard deviation
(right) of the total linear polarization, normalized to the spatially and
temporally averaged intensity, for different integration times (see legend),
for the multiple-slit spectropolarimeter described in the text.
Top panels: without noise.  Bottom panels: with noise.}
\label{F-timeintegralsstats-slit-noscan} 
\end{figure}

For the sake of comparison, we have also computed the variation of the mean
and the standard deviation with time for different integration times
assuming that for each observation we can cover the whole field of view
simultaneously with the necessary number of slits (hereafter, the
multiple-slit spectropolarimeter case; see Fig.
\ref{F-timeintegralsstats-slit-noscan}). By comparing figures
\ref{F-timeintegralsstats}, \ref{F-timeintegralsstats-filter}, and
\ref{F-timeintegralsstats-slit-noscan}, we can conclude that the impact of the time
integration on the characteristics of the polarization signal, at least for
the times relevant to the \ion{Sr}{1} 4607~\AA\ line, are rather reduced.
The largest impact is found for the filter polarimeter case, but this
is purely due to the smaller SNR of the case study, as ignoring the noise
results in a similar decrease in the average and the standard deviation
(see top panels in Fig.  \ref{F-timeintegralsstats-filter}).

\section{Concluding comments}\label{Sconclusions}

We have solved the radiative transfer problem of scattering polarization in the
\ion{Sr}{1} 4607~\AA\ line in a 3D magneto-convection time series model of the
quiet solar photosphere. We studied the impact on the disk-center scattering
polarization map due to the limited time resolution by integrating an increasing
number of snapshots to simulate increasingly longer observing exposures.

As expected, taking into account the finite time resolution of
spectropolarimetric observations in our theoretical results leads to a reduction of
the average linear polarization signal, as well as to a decrease in the visibility
of its spatial variability which we have characterized by means of the standard
deviation.

We chose two instrumental setups of interest, a filter polarimeter similar to
that used by \cite{Zeuneretal2020} and the ViSP instrument at DKIST.

For integration times below the evolution time scale of the granulation
($\lessapprox4$~min), we find that the linear polarization signals in the filter
polarimeter setup are strongly affected by the noise, demonstrated by the
remarkable agreement that can be found between the histograms of the linear
polarization $Q/I$ signals from our simulations and from the unreconstructed
observations in \cite{Zeuneretal2020}. If we instead compare our simulation
without noise with the reconstruction in \cite{Zeuneretal2020}, we can see
that our theoretical forward scattering signals are larger by about a factor $2$.
However, we think that this difference is mainly due to the limitations
of their denoising reconstruction method, and not so much to real intrinsic
differences between the prediction and the observations. It is clear, however,
that further observations with better signal to noise ratio are needed in order
to fully demonstrate this.

We have also shown the type of interesting observations of the disk-center
polarization signals of the \ion{Sr}{1} 4607~\AA\ line that we can expect from
the ViSP instrument at the upcoming DKIST. We must emphasize that, due to the
limited $5$~minutes duration of our time series, we underestimate the effect
of the time evolution of the solar plasma when scanning the field of view, as
we can only loop over the time series, reducing the variability of the
granulation pattern.

In conclusion, our results do indicate that it will be worthwhile to use the
DKIST to map the forward scattering polarization of the \ion{Sr}{1} 4607~\AA\
line to achieve the unprecedented spatio-temporal resolution expected for
such 4~m aperture telescope. This should be attempted using both, a filter
polarimeter having a FWHM significantly smaller than the one considered
in this paper and the ViSP, but the ideal instrument would be a multi-slit
spectropolarimeter capable of simultaneously observing a 2D field of view.
The data that such near-future observations will provide will be very valuable
to probe the unresolved magnetism of the inter-granular lanes via the Hanle
effect.

\acknowledgements

We are grateful to Matthias Rempel (HAO) for having kindly provided the 3D
magneto-convection simulation used in this investigation and to Franziska
Zeuner (IRSOL) for helpful scientific discussions.
Thanks are also due to Roberto Casini (HAO) for clarifying conversations
about the ViSP instrument, and to the referee for useful suggestions after
his/her careful reading of the paper.
We acknowledge the funding received from the European Research Council (ERC)
under the European Union's Horizon 2020 research and innovation programme
(ERC Advanced Grant agreement No 742265).
The 3D radiative transfer simulations were carried out with the MareNostrum
supercomputer of the Barcelona Supercomputing Center (National Supercomputing
Center, Barcelona, Spain), and we gratefully acknowledge the technical
expertise and assistance provided by the Spanish Supercomputing Network,
as well as the additional computer resources used, namely the La Palma
Supercomputer located at the Instituto de Astrof\'isica de Canarias.


\bibliographystyle{apj}
\bibliography{apj-jour,ms.bbl}

\begin{thebibliography}{23}
\expandafter\ifx\csname natexlab\endcsname\relax\def\natexlab#1{#1}\fi

\bibitem[{{Amari} {et~al.}(2015){Amari}, {Luciani}, \& {Aly}}]{Amarietal2015}
{Amari}, T., {Luciani}, J.-F., \& {Aly}, J.-J. 2015, \nat, 522, 188

\bibitem[{{Bianda} {et~al.}(2018){Bianda}, {Berdyugina}, {Gisler}, {Ramelli},
  {Belluzzi}, {Carlin}, {Stenflo}, \& {Berkefeld}}]{Biandaetal2018}
{Bianda}, M., {Berdyugina}, S., {Gisler}, D., {et~al.} 2018, ArXiv e-prints

\bibitem[{{del Pino Alem{\'a}n} {et~al.}(2020){del Pino Alem{\'a}n}, {Trujillo
  Bueno}, {Casini}, \& {Manso Sainz}}]{delPinoetal2020}
{del Pino Alem{\'a}n}, T., {Trujillo Bueno}, J., {Casini}, R., \& {Manso
  Sainz}, R. 2020, \apj, 891, 91

\bibitem[{{del Pino Alem{\'a}n} {et~al.}(2018){del Pino Alem{\'a}n}, {Trujillo
  Bueno}, {{\v S}t{\v e}p{\'a}n}, \& {Shchukina}}]{delPinoetal2018}
{del Pino Alem{\'a}n}, T., {Trujillo Bueno}, J., {{\v S}t{\v e}p{\'a}n}, J., \&
  {Shchukina}, N. 2018, \apj, 863, 164

\bibitem[{{Dhara} {et~al.}(2019){Dhara}, {Capozzi}, {Gisler}, {Bianda},
  {Ramelli}, {Berdyugina}, {Alsina}, \& {Belluzzi}}]{Dharaetal2019}
{Dhara}, S.~K., {Capozzi}, E., {Gisler}, D., {et~al.} 2019, \aap, 630, A67

\bibitem[{{Elmore} {et~al.}(2014){Elmore}, {Rimmele}, {Casini}, {Hegwer},
  {Kuhn}, {Lin}, {McMullin}, {Reardon}, {Schmidt}, {Tritschler}, \&
  {W{\"o}ger}}]{Elmoreetal2014}
{Elmore}, D.~F., {Rimmele}, T., {Casini}, R., {et~al.} 2014, in Society of
  Photo-Optical Instrumentation Engineers (SPIE) Conference Series, Vol. 9147,
  Ground-based and Airborne Instrumentation for Astronomy V (Proc SPIE), 914707

\bibitem[{{Faurobert-Scholl} {et~al.}(1995){Faurobert-Scholl}, {Feautrier},
  {Machefert}, {Petrovay}, \& {Spielfiedel}}]{Faurobertetal1995}
{Faurobert-Scholl}, M., {Feautrier}, N., {Machefert}, F., {Petrovay}, K., \&
  {Spielfiedel}, A. 1995, \aap, 298, 289

\bibitem[{{Fried}(1966)}]{Fried1966}
{Fried}, D.~L. 1966, Journal of the Optical Society of America (1917-1983), 56,
  1372

\bibitem[{{Gandorfer}(2002)}]{BGandorfer2002}
{Gandorfer}, A. 2002, {The Second Solar Spectrum: A high spectral resolution
  polarimetric survey of scattering polarization at the solar limb in graphical
  representation. Volume II: 3910 {\AA} to 4630 {\AA}}

\bibitem[{{Iglesias} {et~al.}(2016){Iglesias}, {Feller}, {Nagaraju}, \&
  {Solanki}}]{Iglesiasetal2016}
{Iglesias}, F.~A., {Feller}, A., {Nagaraju}, K., \& {Solanki}, S.~K. 2016,
  \aap, 590, A89

\bibitem[{{Landi Degl'Innocenti} \& {Landolfi}(2004)}]{BLandiLandolfi2004}
{Landi Degl'Innocenti}, E., \& {Landolfi}, M. 2004, Polarization in Spectral
  Lines (Kluwer Academic Publishers)

\bibitem[{{Manso Sainz} \& {Trujillo Bueno}(2011)}]{MansoTrujillo2011}
{Manso Sainz}, R., \& {Trujillo Bueno}, J. 2011, \apj, 743, 12

\bibitem[{{Rempel}(2014)}]{Rempel2014}
{Rempel}, M. 2014, \apj, 789, 132

\bibitem[{{Rempel}(2017)}]{Rempel2017}
---. 2017, \apj, 834, 10

\bibitem[{{Stenflo} {et~al.}(1997){Stenflo}, {Bianda}, {Keller}, \&
  {Solanki}}]{Stenfloetal1997}
{Stenflo}, J.~O., {Bianda}, M., {Keller}, C.~U., \& {Solanki}, S.~K. 1997,
  \aap, 322, 985

\bibitem[{{Trujillo Bueno}(2001)}]{Trujillo2001}
{Trujillo Bueno}, J. 2001, in Astronomical Society of the Pacific Conference
  Series, Vol. 236, Advanced Solar Polarimetry -- Theory, Observation, and
  Instrumentation, ed. M.~{Sigwarth}, 161

\bibitem[{{Trujillo Bueno} {et~al.}(2002){Trujillo Bueno}, {Landi
  Degl'Innocenti}, {Collados}, {Merenda}, \& {Manso Sainz}}]{Trujilloetal2002}
{Trujillo Bueno}, J., {Landi Degl'Innocenti}, E., {Collados}, M., {Merenda},
  L., \& {Manso Sainz}, R. 2002, \nat, 415, 403

\bibitem[{{Trujillo Bueno} \& {Shchukina}(2007)}]{TrujilloShchukina2007}
{Trujillo Bueno}, J., \& {Shchukina}, N. 2007, \apjl, 664, L135

\bibitem[{{Trujillo Bueno} \& {Shchukina}(2009)}]{TrujilloShchukina2009}
---. 2009, \apj, 694, 1364

\bibitem[{{Trujillo Bueno} {et~al.}(2004){Trujillo Bueno}, {Shchukina}, \&
  {Asensio Ramos}}]{Trujilloetal2004}
{Trujillo Bueno}, J., {Shchukina}, N., \& {Asensio Ramos}, A. 2004, \nat, 430,
  326

\bibitem[{{{\v S}t{\v e}p{\'a}n} \& {Trujillo
  Bueno}(2013)}]{StepanTrujillo2013}
{{\v S}t{\v e}p{\'a}n}, J., \& {Trujillo Bueno}, J. 2013, \aap, 557, A143

\bibitem[{{{\v S}t{\v e}p{\'a}n} \& {Trujillo
  Bueno}(2016)}]{StepanTrujillo2016}
---. 2016, \apjl, 826, L10

\bibitem[{{Zeuner} {et~al.}(2020){Zeuner}, {Manso Sainz}, {Feller}, {van
  Noort}, K., A., {Reardon}, \& {Mart{\'{\i}}nez Pillet}}]{Zeuneretal2020}
{Zeuner}, F., {Manso Sainz}, R., {Feller}, A., {et~al.} 2020, The Astrophysical
  Journal, 893, L44

\end{thebibliography}


\end{document}